\documentclass[twocolumn]{aastex61}
\submitjournal{ApJ}

\shorttitle{Subaru High-$z$ Exploration of Low-Luminosity Quasars (SHELLQs) IV}
\shortauthors{Matsuoka et al.}

\begin{document}

\title{Subaru High-{\scriptsize $z$} Exploration of Low-Luminosity Quasars (SHELLQ{\scriptsize s}). IV. Discovery of 41 Quasars and Luminous Galaxies at $5.7 \le$ {\scriptsize $z$} $\le 6.9$}

\correspondingauthor{Yoshiki Matsuoka}
\email{yk.matsuoka@cosmos.ehime-u.ac.jp}

\author{Yoshiki Matsuoka}
\affil{Research Center for Space and Cosmic Evolution, Ehime University, Matsuyama, Ehime 790-8577, Japan.}

\author{Kazushi Iwasawa}
\affil{ICREA and Institut de Ci{\`e}ncies del Cosmos, Universitat de Barcelona, IEEC-UB, Mart{\'i} i Franqu{\`e}s, 1, 08028 Barcelona, Spain.}

\author{Masafusa Onoue}
\affil{National Astronomical Observatory of Japan, Mitaka, Tokyo 181-8588, Japan.}
\affil{Department of Astronomical Science, Graduate University for Advanced Studies (SOKENDAI), Mitaka, Tokyo 181-8588, Japan.}

\author{Nobunari Kashikawa}
\affil{National Astronomical Observatory of Japan, Mitaka, Tokyo 181-8588, Japan.}
\affil{Department of Astronomical Science, Graduate University for Advanced Studies (SOKENDAI), Mitaka, Tokyo 181-8588, Japan.}

\author{Michael A. Strauss}
\affil{Princeton University Observatory, Peyton Hall, Princeton, NJ 08544, USA.}

\author{Chien-Hsiu Lee}
\affil{Subaru Telescope, Hilo, HI 96720, USA.}

\author{Masatoshi Imanishi}
\affil{National Astronomical Observatory of Japan, Mitaka, Tokyo 181-8588, Japan.}
\affil{Department of Astronomical Science, Graduate University for Advanced Studies (SOKENDAI), Mitaka, Tokyo 181-8588, Japan.}

\author{Tohru Nagao}
\affil{Research Center for Space and Cosmic Evolution, Ehime University, Matsuyama, Ehime 790-8577, Japan.}

\author{Masayuki Akiyama}
\affil{Astronomical Institute, Tohoku University, Aoba, Sendai, 980-8578, Japan.}

\author{Naoko Asami}
\affil{Seisa University, Hakone-machi, Kanagawa, 250-0631, Japan.}

\author{James Bosch}
\affil{Princeton University Observatory, Peyton Hall, Princeton, NJ 08544, USA.}


\author{Hisanori Furusawa}
\affil{National Astronomical Observatory of Japan, Mitaka, Tokyo 181-8588, Japan.}

\author{Tomotsugu Goto}
\affil{Institute of Astronomy and Department of Physics, National Tsing Hua University, Hsinchu 30013, Taiwan.}

\author{James E. Gunn}
\affil{Princeton University Observatory, Peyton Hall, Princeton, NJ 08544, USA.}

\author{Yuichi Harikane}
\affil{Institute for Cosmic Ray Research, The University of Tokyo, Kashiwa, Chiba 277-8582, Japan}
\affil{Department of Physics, Graduate School of Science, The University of Tokyo, Bunkyo, Tokyo 113-0033, Japan}

\author{Hiroyuki Ikeda}
\affil{National Astronomical Observatory of Japan, Mitaka, Tokyo 181-8588, Japan.}

\author{Takuma Izumi}
\affil{National Astronomical Observatory of Japan, Mitaka, Tokyo 181-8588, Japan.}

\author{Toshihiro Kawaguchi}
\affil{Department of Economics, Management and Information Science, Onomichi City University, Onomichi, Hiroshima 722-8506, Japan.}

\author{Nanako Kato}
\affil{Graduate School of Science and Engineering, Ehime University, Matsuyama, Ehime 790-8577, Japan.}

\author{Satoshi Kikuta}
\affil{National Astronomical Observatory of Japan, Mitaka, Tokyo 181-8588, Japan.}
\affil{Department of Astronomical Science, Graduate University for Advanced Studies (SOKENDAI), Mitaka, Tokyo 181-8588, Japan.}

\author{Kotaro Kohno}
\affil{Institute of Astronomy, The University of Tokyo, Mitaka, Tokyo 181-0015, Japan.}
\affil{Research Center for the Early Universe, University of Tokyo, Tokyo 113-0033, Japan.}

\author{Yutaka Komiyama}
\affil{National Astronomical Observatory of Japan, Mitaka, Tokyo 181-8588, Japan.}
\affil{Department of Astronomical Science, Graduate University for Advanced Studies (SOKENDAI), Mitaka, Tokyo 181-8588, Japan.}

\author{Robert H. Lupton}
\affil{Princeton University Observatory, Peyton Hall, Princeton, NJ 08544, USA.}

\author{Takeo Minezaki}
\affil{Institute of Astronomy, The University of Tokyo, Mitaka, Tokyo 181-0015, Japan.}

\author{Satoshi Miyazaki}
\affil{National Astronomical Observatory of Japan, Mitaka, Tokyo 181-8588, Japan.}
\affil{Department of Astronomical Science, Graduate University for Advanced Studies (SOKENDAI), Mitaka, Tokyo 181-8588, Japan.}

\author{Tomoki Morokuma}
\affil{Institute of Astronomy, The University of Tokyo, Mitaka, Tokyo 181-0015, Japan.}

\author{Hitoshi Murayama}
\affil{Kavli Institute for the Physics and Mathematics of the Universe, WPI, The University of Tokyo,Kashiwa, Chiba 277-8583, Japan.}

\author{Mana Niida}
\affil{Graduate School of Science and Engineering, Ehime University, Matsuyama, Ehime 790-8577, Japan.}

\author{Atsushi J. Nishizawa}
\affil{Institute for Advanced Research, Nagoya University, Furo-cho, Chikusa-ku, Nagoya 464-8602, Japan.}

\author{Masamune Oguri}
\affil{Department of Physics, Graduate School of Science, The University of Tokyo, Bunkyo, Tokyo 113-0033, Japan}
\affil{Kavli Institute for the Physics and Mathematics of the Universe, WPI, The University of Tokyo,Kashiwa, Chiba 277-8583, Japan.}
\affil{Research Center for the Early Universe, University of Tokyo, Tokyo 113-0033, Japan.}

\author{Yoshiaki Ono}
\affil{Institute for Cosmic Ray Research, The University of Tokyo, Kashiwa, Chiba 277-8582, Japan}

\author{Masami Ouchi}
\affil{Institute for Cosmic Ray Research, The University of Tokyo, Kashiwa, Chiba 277-8582, Japan}
\affil{Kavli Institute for the Physics and Mathematics of the Universe, WPI, The University of Tokyo,Kashiwa, Chiba 277-8583, Japan.}

\author{Paul A. Price}
\affil{Princeton University Observatory, Peyton Hall, Princeton, NJ 08544, USA.}

\author{Hiroaki Sameshima}
\affil{Koyama Astronomical Observatory, Kyoto-Sangyo University, Kita, Kyoto, 603-8555, Japan.}

\author{Andreas Schulze}
\affil{National Astronomical Observatory of Japan, Mitaka, Tokyo 181-8588, Japan.}

\author{Hikari Shirakata}
\affil{Department of Cosmosciences, Graduates School of Science, Hokkaido University, Kitaku, Sapporo 060-0810, Japan.}

\author{John D. Silverman}
\affil{Kavli Institute for the Physics and Mathematics of the Universe, WPI, The University of Tokyo, Kashiwa, Chiba 277-8583, Japan.}

\author{Naoshi Sugiyama}
\affil{Kavli Institute for the Physics and Mathematics of the Universe, WPI, The University of Tokyo, Kashiwa, Chiba 277-8583, Japan.}
\affil{Graduate School of Science, Nagoya University, Furo-cho, Chikusa-ku, Nagoya 464-8602, Japan.}

\author{Philip J. Tait}
\affil{Subaru Telescope, Hilo, HI 96720, USA.}

\author{Masahiro Takada}
\affil{Kavli Institute for the Physics and Mathematics of the Universe, WPI, The University of Tokyo,Kashiwa, Chiba 277-8583, Japan.}

\author{Tadafumi Takata}
\affil{National Astronomical Observatory of Japan, Mitaka, Tokyo 181-8588, Japan.}
\affil{Department of Astronomical Science, Graduate University for Advanced Studies (SOKENDAI), Mitaka, Tokyo 181-8588, Japan.}

\author{Masayuki Tanaka}
\affil{National Astronomical Observatory of Japan, Mitaka, Tokyo 181-8588, Japan.}
\affil{Department of Astronomical Science, Graduate University for Advanced Studies (SOKENDAI), Mitaka, Tokyo 181-8588, Japan.}

\author{Ji-Jia Tang}
\affil{Institute of Astronomy and Astrophysics, Academia Sinica, Taipei, 10617, Taiwan.}

\author{Yoshiki Toba}
\affil{Research Center for Space and Cosmic Evolution, Ehime University, Matsuyama, Ehime 790-8577, Japan.}
\affil{Institute of Astronomy and Astrophysics, Academia Sinica, Taipei, 10617, Taiwan.}
\affil{Department of Astronomy, Kyoto University, Kitashirakawa-Oiwake-cho, Sakyo-ku, Kyoto 606-8502, Japan.}

\author{Yousuke Utsumi}
\affil{Kavli Institute for Particle Astrophysics and Cosmology, Stanford University, CA 94025, USA.}

\author{Shiang-Yu Wang}
\affil{Institute of Astronomy and Astrophysics, Academia Sinica, Taipei, 10617, Taiwan.}

\author{Takuji Yamashita}
\affil{Research Center for Space and Cosmic Evolution, Ehime University, Matsuyama, Ehime 790-8577, Japan.}



\begin{abstract}

We report discovery of 41 new high-$z$ quasars and luminous galaxies, which were spectroscopically identified at $5.7 \le z \le 6.9$.
This is the fourth in a series of papers from the Subaru High-$z$ Exploration of Low-Luminosity Quasars (SHELLQs) project, based on 
the deep multi-band imaging data collected by the Hyper Suprime-Cam (HSC) Subaru Strategic Program survey.
We selected the photometric candidates by a Bayesian probabilistic algorithm, and then carried out follow-up spectroscopy with the Gran Telescopio Canarias and the Subaru Telescope.
Combined with the sample presented in the previous papers, we have now spectroscopically identified 137 extremely-red HSC sources over about 650 deg$^2$, which include
64 high-$z$ quasars, 24 high-$z$ luminous galaxies, 6 [\ion{O}{3}] emitters at $z \sim 0.8$, and 43 Galactic cool dwarfs (low-mass stars and brown dwarfs).
The new quasars span the luminosity range from $M_{1450} \sim -26$ to $-22$ mag, and continue to populate a few magnitude lower luminosities 
than have been probed by previous wide-field surveys.
In a companion paper, we derive the quasar luminosity function at $z \sim 6$ over an unprecedentedly wide range of $M_{1450} \sim -28$ to $-21$ mag, 
exploiting the SHELLQs and other survey outcomes. 
\end{abstract}




\section{Introduction} \label{sec:intro}

High-$z$ quasars\footnote{Throughout this paper, ``high-$z$" denotes $z > 5.7$, where quasars are observed as $i$-band dropouts in the 
Sloan Digital Sky Survey (SDSS) filter system \citep{fukugita96}.
The term ``$X$-band dropouts" or ``$X$-dropouts" refers to objects which are much fainter in the $X$ and bluer bands (and very often invisible) than in the redder bands.} 
are an important probe of the early Universe in many aspects. 
Their rest-ultraviolet (UV) spectra bluewards of Ly$\alpha$ are very sensitive to the \ion{H}{1} absorption, and thus indicate the progress of cosmic reionization through the neutral fraction of the intergalactic medium 
\citep[IGM;][]{gunn65, fan06araa}.
The very existence of such objects puts strong restriction on, and sometimes even challenges, models of formation and early evolution of supermassive black holes
\citep[SMBH; e.g.,][]{volonteri12, ferrara14, madau14}.
High-$z$ quasars are also used to probe the assembly of the host galaxies, which are thought to form
in the highest density peaks of the dark matter distribution in the early phase of cosmic history \citep[e.g.,][]{goto09, decarli17}.

Previous surveys have identified more than 100 high-$z$ quasars so far \citep[e.g.,][and references therein]{banados16}, with the most distant 
objects known at $z = 7.54$ \citep{banados18} and $z = 7.085$ \citep{mortlock11}.
However, most of the known quasars have redshifts $z < 6.5$ and UV absolute magnitudes $M_{1450} < -24$ mag, while higher redshifts 
and fainter magnitudes remain largely unexplored.
There must be numerous objects of  faint quasars and active galactic nuclei (AGNs) behind the known luminous quasars; they 
may represent the more typical mode of SMBH growth in the early Universe, and may have made a significant contribution to reionization.

For the past few years, we have been carrying out a high-$z$ quasar survey with the Subaru 8.2-m telescope.
We have already reported discovery of 33 high-$z$ quasars, along with 14 high-$z$ luminous galaxies, 2 [\ion{O}{3}] emitters at $z \sim 0.8$, and 15 Galactic
brown dwarfs, in \citet[][Paper I and II, hereafter]{matsuoka16,matsuoka17}.
Multi-wavelength follow-up observations of the discovered objects are ongoing, whose initial results from the Atacama Large Millimeter/ submillimeter Array (ALMA) data are presented 
in \citet[][Paper III]{izumi18}.
This ``Subaru High-$z$ Exploration of Low-Luminosity Quasars (SHELLQs)'' project is  based on the exquisite multi-band photometry data collected by 
the Hyper Suprime-Cam (HSC) Subaru Strategic Program (SSP) survey \citep{aihara17_survey}.
HSC is a wide-field camera on the Subaru Telescope, and has a nearly circular field of view of 1$^\circ$.5 diameter, covered 
by 116 2K $\times$ 4K Hamamatsu fully depleted CCDs, with the pixel scale of 0\arcsec.17 \citep{miyazaki18}.
The camera dewar design and on-site quality assurance design are described in \citet{komiyama18} and \citet{furusawa18}, respectively.
The HSC-SSP survey has three layers with different combinations of area and depth.
The Wide layer is observing 1400 deg$^2$ mostly along the celestial equator, with the 
5$\sigma$ point-source depths of ($g_{\rm AB}$, $r_{\rm AB}$, $i_{\rm AB}$, $z_{\rm AB}$, $y_{\rm AB}$) = (26.5, 26.1, 25.9, 25.1, 24.4) mag 
measured in 2\arcsec.0 aperture.
The Deep and the UltraDeep layers are observing smaller areas (27 and 3.5 deg$^2$) down to deeper limiting magnitudes ($r_{\rm AB}$ = 27.1 and 27.7 mag, respectively).
The observed data are processed 
with the dedicated pipeline {\it hscPipe} \citep{bosch18}, which was developed from the Large Synoptic Survey Telescope software pipeline \citep{juric15}.
The HSC pipeline is evolving through the period of the survey, which sometimes gives rise to somewhat inconsistent photometry (and quasar probability of a given source; see below) 
between different data releases.
A full description of the HSC-SSP survey may be found in \citet{aihara17_survey}. 

This paper is the fourth in a series of SHELLQs publications, and reports spectroscopic identification of an additional 73 objects
in the latest HSC-SSP data. 
We describe the photometric candidate selection briefly in \S \ref{sec:selection}, while a more complete description may be found in Paper I and II.
The spectroscopic follow-up observations are described in \S \ref{sec:spectroscopy}.
The quasars and other classes of objects we discovered are presented in \S \ref{sec:results}.
A summary appears in \S \ref{sec:summary}.
A companion paper (Y. Matsuoka et al., in prep.) describes the quasar luminosity function at $z \sim 6$ derived from the SHELLQs sample obtained so far.
This paper adopts the cosmological parameters $H_0$ = 70 km s$^{-1}$ Mpc$^{-1}$, $\Omega_{\rm M}$ = 0.3, and $\Omega_{\rm \Lambda}$ = 0.7.
All magnitudes in the optical and NIR bands are presented in the AB system \citep{oke83}, and are corrected for Galactic extinction \citep{schlegel98}.
We use two types of magnitudes; the point spread function (PSF) magnitude ($m_{\rm AB}$) is measured by fitting PSF models to the source profile,
while the CModel magnitude ($m_{\rm CModel, AB}$) is measured by fitting PSF-convolved, two-component galaxy models \citep{abazajian04}.
We use the PSF magnitude error ($\sigma_{\rm m}$) measured by the HSC data processing pipeline \citep{bosch18}; it does not contain photometric calibration uncertainty, 
which is estimated to be at least 1 \% \citep{aihara17_pdr1}.
In what follows, we refer to $z$-band magnitudes with the AB subscript (``$z_{\rm AB}$"), while redshift $z$ appears without a subscript.

\section{Photometric Candidate Selection} \label{sec:selection}

Our photometric candidate selection starts from the HSC-SSP source catalog. 
We used the survey data in the three layers, observed before 2017 May, i.e., a newer dataset than contained in the latest public data release \citep{aihara17_pdr1}.
We require that our quasar survey field has been observed in the $i$, $z$, and $y$ bands (not necessarily to the full depths; see below), 
but impose no requirement on  the $g$- or $r$-band coverage.
All sources meeting the following criteria, without critical quality flags (see Paper I), are selected:
\begin{eqnarray}
(z_{\rm AB} < 24.5\ {\rm and}\ \sigma_z < 0.155\ {\rm and}\ i_{\rm AB} - z_{\rm AB} > 1.5 \nonumber \\
 {\rm and}\ z_{\rm AB} - z_{\rm CModel, AB} < 0.15 )
\label{eq:query1}
\end{eqnarray}
or
\begin{eqnarray}
(y_{\rm AB} < 24.0\ {\rm and}\ \sigma_y < 0.155\ {\rm and}\ z_{\rm AB} - y_{\rm AB} > 0.8 \nonumber \\
 {\rm and}\ y_{\rm AB} - y_{\rm CModel, AB} < 0.15 ) .
\label{eq:query2}
\end{eqnarray}
Here we use the difference between the PSF and CModel magnitudes ($m_{\rm AB} - m_{\rm CModel, AB}$) to reject extended sources.
With the present threshold value (0.15), our completeness of point source selection is $>$80 \% at $z_{\rm AB} < 24.0$ mag (Paper II).
The completeness decreases toward fainter magnitude, which should be accounted for in statistical measurements (e.g., luminosity function) of the discovered quasars.
We further remove sources with more than $3\sigma$ detection in the $g$ or $r$ band, if these bands are available, as such sources are most likely low-$z$ interlopers.


Next, we process the HSC images of the above sources through 
an automatic image checking procedure, 
in all the available bands. 
As detailed in Paper I and II, this procedure uses Source Extractor \citep{bertin96}, and removes those sources whose photometry is not consistent 
(within 5$\sigma$ significance) between the stacked and all the pre-stacked, individual images. 
We also eliminate sources with too compact, diffuse, or elliptical profiles to be celestial point sources 
on the stacked images.
The sources removed at this stage are mostly cosmic rays, moving or transient sources, or image artifacts.

The candidates selected above are matched, within 1\arcsec.0 separation, to the public near-IR catalogs from
the United Kingdom Infrared Telescope Infrared Deep Sky Survey \citep{lawrence07},  
the Visible and Infrared Survey Telescope for Astronomy (VISTA) Kilo-degree Infrared Galaxy survey, the VISTA Deep Extragalactic Observations Survey \citep{jarvis13},
and the UltraVISTA survey \citep{mccracken12}.
The choice of the 1\arcsec.0 matching radius is rather arbitrary, but is at least sufficiently larger than the astrometric uncertainties of the above surveys.
Then using the $i$, $z$ and $y$-band magnitudes along with all the available $J$, $H$, and/or $K$ magnitudes, we calculate Bayesian quasar probability ($P_{\rm Q}^{\rm B}$) for each candidate,
based on spectral energy distribution (SED) models and estimated surface densities of high-$z$ quasars and contaminating brown dwarfs as a function of magnitude. 
Our algorithm does not contain galaxy models 
at present.
We created the quasar SED models 
by stacking the SDSS spectra of 340 bright quasars at $z \simeq 3$, where the quasar color selection is 
fairly complete \citep{richards02, willott05}, and correcting for the effect of IGM absorption \citep{songaila04}.
The quasar surface density was modeled with the luminosity function of \citet{willott10}.
The color models of brown dwarfs were computed with a set of observed spectra compiled in the SpeX prism library\footnote{\url{http://pono.ucsd.edu/~adam/ browndwarfs/spexprism}} 
and the CGS4 library\footnote{\url{http://staff.gemini.edu/~sleggett/LTdata.html}}, while the surface densities were calculated following \citet{caballero08}.
A more detailed description of our Bayesian algorithm may be found in Paper I.
We reject sources with $P_{\rm Q}^{\rm B} < 0.1$, and keep those with higher quasar probability in the sample of candidates.


Finally, we inspect the images of all the candidates by eye and remove additional problematic sources.
About 80 \% of the remaining candidates were rejected at this last stage, which are mostly cosmic rays but also include transient/moving objects and sources close to bright stars.  
The latest HSC-SSP (internal) data release covers 650 deg$^2$, when we limit to the field where at least a single exposure in the $i$ and two exposures each in the $z$ and $y$ bands
were obtained.
From the final sample of candidates, 
we put the highest priority for spectroscopic identification on a subsample 
with relatively bright magnitudes ($z_{\rm AB} <$ 24.0 mag or $y_{\rm AB} <$ 23.5 mag), red colors ($i_{\rm AB} - z_{\rm AB} > 2.0$ or $z_{\rm AB} - y_{\rm AB} > 0.8$),
and detection in more than a single band (for $i$-dropouts) 
or a single exposure (for $z$-dropouts).
The number of our photometric candidates is changing throughout the survey, due to continuous arrival of new HSC-SSP survey data, a change in the photometry and quasar probability ($P_{\rm Q}^{\rm B}$) with updates of the data processing pipeline,
and the progress of our follow-up spectroscopy.

The present HSC survey footprint includes 15 high-$z$ quasars discovered by other surveys, as summarized in Table \ref{tab:recovery}. 
We recovered ten of these quasars with $P_{\rm Q}^{\rm B} = 1.000$, while three quasars were not selected due to their relatively low redshifts 
($z < 5.9$) and bluer HSC colors ($i_{\rm AB} - z_{\rm AB} < 1.5$) than the HSC color selection threshold described above.
The remaining two quasars are missing due to nearby bright sources; VIK $J0839+0015$ is assigned with an HSC {\tt pixelflags\_bright\_objectcenter} flag (indicating that the pipeline measurements of this source may be affected by a nearby bright source)
and therefore does not meet the HSC source selection criteria described above, while SDSS $J1602+4228$ is rejected in our automatic image checking procedure, since its $i$-band photometry is not consistent between stacked and pre-stacked images, 
likely due to blending with a foreground galaxy.

\begin{deluxetable*}{cccccl}
\tablecaption{Recovery of Known High-$z$ Quasars \label{tab:recovery}}
\tablehead{
\colhead{Name} &\colhead{R.A.} &\colhead{Decl.} &  \colhead{Redshift} & \colhead{Recovered?} & 
\colhead{Comment} 
} 
\startdata
  CFHQS $J0210-0456$         & 02$^{\rm h}$10$^{\rm m}$13$^{\rm s}$19  & $-$04$^\circ$56\arcmin20\arcsec.9  & 6.43 & Y & $P_{\rm Q}^{\rm B} = 1.000$\\
  CFHQS $J0216-0455$         & 02$^{\rm h}$16$^{\rm m}$27$^{\rm s}$.81  & $-$04$^\circ$55\arcmin34\arcsec.1  & 6.01 & Y & $P_{\rm Q}^{\rm B} = 1.000$\\
  CFHQS $J0227-0605$         & 02$^{\rm h}$27$^{\rm m}$43$^{\rm s}$.29  & $-$06$^\circ$05\arcmin30\arcsec.2  & 6.20 & Y & $P_{\rm Q}^{\rm B} = 1.000$\\
  SDSS $J0836+0054$           & 08$^{\rm h}$36$^{\rm m}$43$^{\rm s}$.86 & $+$00$^\circ$54\arcmin53\arcsec.3 & 5.81 & N & $i_{\rm AB} - z_{\rm AB} < 1.5$\\
  VIK $J0839+0015$               & 08$^{\rm h}$39$^{\rm m}$55$^{\rm s}$.36   & $+$00$^\circ$15\arcmin54\arcsec.2 & 5.84 & N & Close to a bright source\\
  VIK $J1148+0056$               &11$^{\rm h}$48$^{\rm m}$33$^{\rm s}$.18   & $+$00$^\circ$56\arcmin42\arcsec.3 & 5.84 & N & $i_{\rm AB} - z_{\rm AB} < 1.5$\\
  VIK $J1215+0023$               &12$^{\rm h}$15$^{\rm m}$16$^{\rm s}$.87   & $+$00$^\circ$23\arcmin24\arcsec.7 & 5.93 & Y & $P_{\rm Q}^{\rm B} = 1.000$\\
  PSO $J184.3389+01.5284$ & 12$^{\rm h}$17$^{\rm m}$21$^{\rm s}$.34  & $+$01$^\circ$31\arcmin42\arcsec.5 & 6.20 & Y & $P_{\rm Q}^{\rm B} = 1.000$\\
  SDSS $J1602+4228$           &16$^{\rm h}$02$^{\rm m}$53$^{\rm s}$.98   & $+$42$^\circ$28\arcmin24\arcsec.9 & 6.09 & N & Blended with a foreground galaxy\\
  IMS $J2204+0012$              & 22$^{\rm h}$04$^{\rm m}$17$^{\rm s}$.92   & $+$01$^\circ$11\arcmin44\arcsec.8 & 5.94 & Y & $P_{\rm Q}^{\rm B} = 1.000$\\
  VIMOS 2911001793             & 22$^{\rm h}$19$^{\rm m}$17$^{\rm s}$.22   & $+$01$^\circ$02\arcmin48\arcsec.9 & 6.16 & Y & $P_{\rm Q}^{\rm B} = 1.000$\\
  SDSS $J2228+0110$           & 22$^{\rm h}$28$^{\rm m}$43$^{\rm s}$.54  & $+$01$^\circ$10\arcmin32\arcsec.2 & 5.95 & Y & $P_{\rm Q}^{\rm B} = 1.000$\\
  CFHQS $J2242+0334$        & 22$^{\rm h}$42$^{\rm m}$37$^{\rm s}$.55 & $+$03$^\circ$34\arcmin21\arcsec.6 & 5.88 & N & $i_{\rm AB} - z_{\rm AB} < 1.5$\\
  SDSS $J2307+0031$           & 23$^{\rm h}$07$^{\rm m}$35$^{\rm s}$.35  & $+$00$^\circ$31\arcmin49\arcsec.4 & 5.87 & Y & $P_{\rm Q}^{\rm B} = 1.000$\\
  SDSS $J2315-0023$            & 23$^{\rm h}$15$^{\rm m}$46$^{\rm s}$.57  & $-$00$^\circ$23\arcmin58\arcsec.1  & 6.12 & Y & $P_{\rm Q}^{\rm B} = 1.000$\\
\enddata
\tablecomments{The naming convention follows \citet{banados16}.}
\end{deluxetable*}

\section{Spectroscopy} \label{sec:spectroscopy}

Our previous papers reported the results of follow-up spectroscopy carried out before 2016 Autumn.
Since that time, we have observed 73 additional quasar candidates,
using the Optical System for Imaging and low-Intermediate-Resolution Integrated Spectroscopy \citep[OSIRIS;][]{cepa00} mounted on the 10.4-m Gran Telescopio Canarias (GTC),
and the Faint Object Camera and Spectrograph \citep[FOCAS;][]{kashikawa02} mounted on Subaru.
We observed roughly the brightest one-third of the candidates with OSIRIS, and the remaining candidates with FOCAS.
The observations were scheduled 
in such a way that the targets with brighter magnitudes and higher $P_{\rm Q}^{\rm B}$ were observed at the earlier opportunities.
The journal of these discovery observations is presented in Table \ref{tab:obsjournal}.
We present the details of these observations below.

\subsection{GTC/OSIRIS}

GTC is a 10.4-m telescope located at the Observatorio del Roque de los Muchachos in La Palma, Spain.
Our program was awarded 26 hours in the 2017A semester (GTC8-17A; Iwasawa et al.).
We used OSIRIS with the R2500I grism and 1\arcsec.0-wide long slit, which provides spectral coverage from $\lambda_{\rm obs}$ = 0.74 to 1.0\ $\mu$m 
with a resolution $R \sim 1500$.
The observations were carried out in queue mode on dark and gray nights, with mostly photometric (sometimes spectroscopic) sky conditions and the seeing 0\arcsec.7 -- 1\arcsec.3.
The data were reduced using the Image Reduction and Analysis Facility (IRAF).
Bias correction, flat fielding with dome flats, sky subtraction, and 1d extraction were performed in the standard way.
The wavelength calibration was performed with reference to sky emission lines.
The flux calibration was tied to white dwarfs (Ross 640, Feige 110, G191-B2B) or a B-type standard star (HILT 600), observed as standard stars within a few days of the target observations.
The slit losses were corrected for by scaling the spectra to match the HSC magnitudes in the $z$ and $y$ bands for the $i$- and $z$-band dropouts, respectively.


\subsection{Subaru/FOCAS}

Our program was awarded five nights each in the S17A and S17B semesters, as a part of a Subaru intensive program (S16B-071I; Matsuoka et al.).
We used FOCAS in the multi-object spectrograph (MOS) mode with the VPH900 grism and SO58 order-sorting filter.
The widths of the slitlets were set to 1\arcsec.0.
This configuration provides spectral coverage from $\lambda_{\rm obs}$ = 0.75 to 1.05\ $\mu$m with a resolution $R \sim 1200$.
The only exception is $J1400+0106$, which was observed with a 2\arcsec.0-wide longslit with the same grism and filter as the MOS observations.
All the observations were carried out on gray nights.
A few of these nights were occasionally affected by cirrus or poor seeing, while the weather was fairly good
with seeing 0\arcsec.4 -- 1\arcsec.2 for the rest of the observations.

The data were reduced with IRAF using the dedicated FOCASRED package.
Bias correction, flat fielding with dome flats, sky subtraction, and 1d extraction were performed in the standard way.
The wavelength was calibrated with reference to the sky emission lines.
The flux calibration was tied to white dwarf standard stars (G191-B2B, Feige 110, or GD 153) observed during the same run, in most cases on the same nights as the targets.
The slit losses were corrected for 
in the same way as in the OSIRIS data reductions.

\clearpage
\startlongtable
\begin{deluxetable*}{cccc|cccc|cccc}
\tablecaption{Journal of Discovery Spectroscopy\label{tab:obsjournal}}
\tablehead{
\colhead{Target} & \colhead{$t_{\rm exp}$} & 
\colhead{Date} & \colhead{Inst} & 
\colhead{Target} & \colhead{$t_{\rm exp}$} & 
\colhead{Date} & \colhead{Inst} &
\colhead{Target} & \colhead{$t_{\rm exp}$} & 
\colhead{Date} & \colhead{Inst}\\
\colhead{} & \colhead{(min)} & 
\colhead{} & \colhead{} & 
\colhead{} & \colhead{(min)} & 
\colhead{} & \colhead{} &
\colhead{} & \colhead{(min)} & 
\colhead{} & \colhead{}
} 
\startdata
  $J2210+0304$ & 210 & Sep 29, Oct 1 & F & $J0858+0000$ & 15 & Mar 1 & O              & $J0225-0351$ & 23 & Oct 1 & F\\
  $J0213-0626$  & 80 & Sep 28, Oct 1  & F  & $J0220-0432$ & 70 & Sep 30 & F             & $J0837-0000$ & 15 & Apr 23 & O   \\
  $J0923+0402$ & 30 & Mar 2 & O                & $J1422+0011$ & 30 & May 3 & F              & $J0856+0248$ & 30 & Mar 19 & F  \\
  $J0921+0007$ & 15 & Mar 1 & O                & $J2231-0035$ & 73 & Sep 30 & F             & $J0900+0424$ & 40 & Mar 18 & F    \\
  $J1545+4232$ & 30 & Mar 7 & O                & $J1209-0006$ & 60 & Mar 17 & F             & $J0902-0030$ & 30 & Mar 18 & F   \\
  $J1004+0239$ & 30 & Mar 18 & F               & $J1550+4318$ & 45 & Mar 17 & F            & $J0906-0206$ & 45 & Mar 1& O    \\
  $J0211-0203$ & 110 & Sep 27, Oct 1 & F   & $J1428+0159$ & 30 & Feb 15 (2018) & O            & $J0906+0431$ & 45 & Mar 18 & F   \\
  $J2304+0045$ & 30 & Sep 27 & F              & $J0917-0056$ & 120 & Mar 2, Apr 23 & O & $J0912-0121$ & 30 & Mar 16 & F \\
  $J2255+0251$ & 90 & Sep 27, Oct 1 & F   & $J0212-0315$ & 50 & Sep 27 & F               & $J1359+0134$ & 30 & Mar 18 & F \\
  $J1406-0116$ & 30 & Mar 3 & O                 & $J0212-0532$ & 50 & Sep 29 & F               & $J1400+0106$ & 240 & Mar 19 & F   \\
  $J1146-0005$ & 60 & Mar 7 & O                 & $J2311-0050$ & 50 & Sep 30 & F              & $J1415-0113$ & 15 & Apr 23 & O  \\
  $J1146+0124$ & 45 & Mar 29 & O              & $J1609+5515$ & 80 & Mar 17 & F              & $J1432+0045$ & 75 & Mar 18 & F\\
  $J0918+0139$ & 60 & Mar 2 & O                & $J1006+0300$ & 45 & Mar 16 & F              & $J1434-0204$ & 15 & Apr 22 & O  \\
  $J0844-0132$ & 60 & Dec 10 & O               & $J0914+0442$ & 50 & Mar 19 & F              & $J1435+0040$ & 15 & Apr 22 & O\\
  $J1146-0154$ & 60 & Mar 31 & O                & $J0219-0132$ & 30 & Sep 28 & F              & $J1607+5417$ & 30 & Mar 18 & F \\
  $J0834+0211$ & 40 & Dec 10 & O               & $J0915-0051$ & 75 & Mar 18 & F              & $J1620+4438$ & 20 & Mar 18 & F   \\
  $J0909+0440$ & 15 & Mar 1 & O                & $J1000+0211$ & 13 & Mar 17 & F               & $J1629+4233$ & 30 & May 3 & F\\
  $J2252+0225$ & 100 & Sep 29, Oct 1 & F  & $J0154-0116$ & 14 & Sep 28 & F               & $J2236+0006$ & 50 & Sep 29 & F \\
  $J1406-0144$ & 60 & Apr 22 & O                & $J2226+0237$ & 10 & Sep 30 & F              & $J2239-0048$ & 25 & Oct 1 & F \\
  $J1416+0147$ & 30 & Apr 22 & O               & $J0845-0123$ & 10 & Mar 16 & F               & $J2248+0103$ & 50 &Sep 29 & F  \\
  $J0957+0053$ & 75 & Mar 16 & F               & $J0158-0033$ & 30 & Sep 30 & F              & $J2253-0117$ & 40 & Dec 14 & O \\
  $J2223+0326$ & 30 & Sep 27 & F               & $J0207-0052$ & 10 & Sep 27 & F              & $J2305-0051$ & 80 & Sep 28 & F \\
  $J1400-0125$ & 50 & Mar 16 & F                & $J0213-0334$ & 10 & Sep 30 & F              & $J2315-0041$ & 25 & Dec 22 & O  \\
  $J1400-0011$ & 80 & Mar 6, Apr 23 & O     & $J0216-0207$ &  30 & Sep 30 & F             &                         &       &             &  \\
$J1219+0050$ & 30 & Dec 22 & O                & $J0220-0134$ & 30 & Sep 28 & F               &                         &       &             &  \\
\enddata
\tablecomments{All the dates are in the year of 2017, except for $J1428+0159$ observed in 2018. 
The instrument (Inst) ``O" and ``F" denote GTC/OSIRIS and Subaru/FOCAS, respectively.}
\end{deluxetable*}

\ {\ }
\clearpage
\section{Results and Discussion \label{sec:results}}

\begin{figure*}
\epsscale{1.0}
\plotone{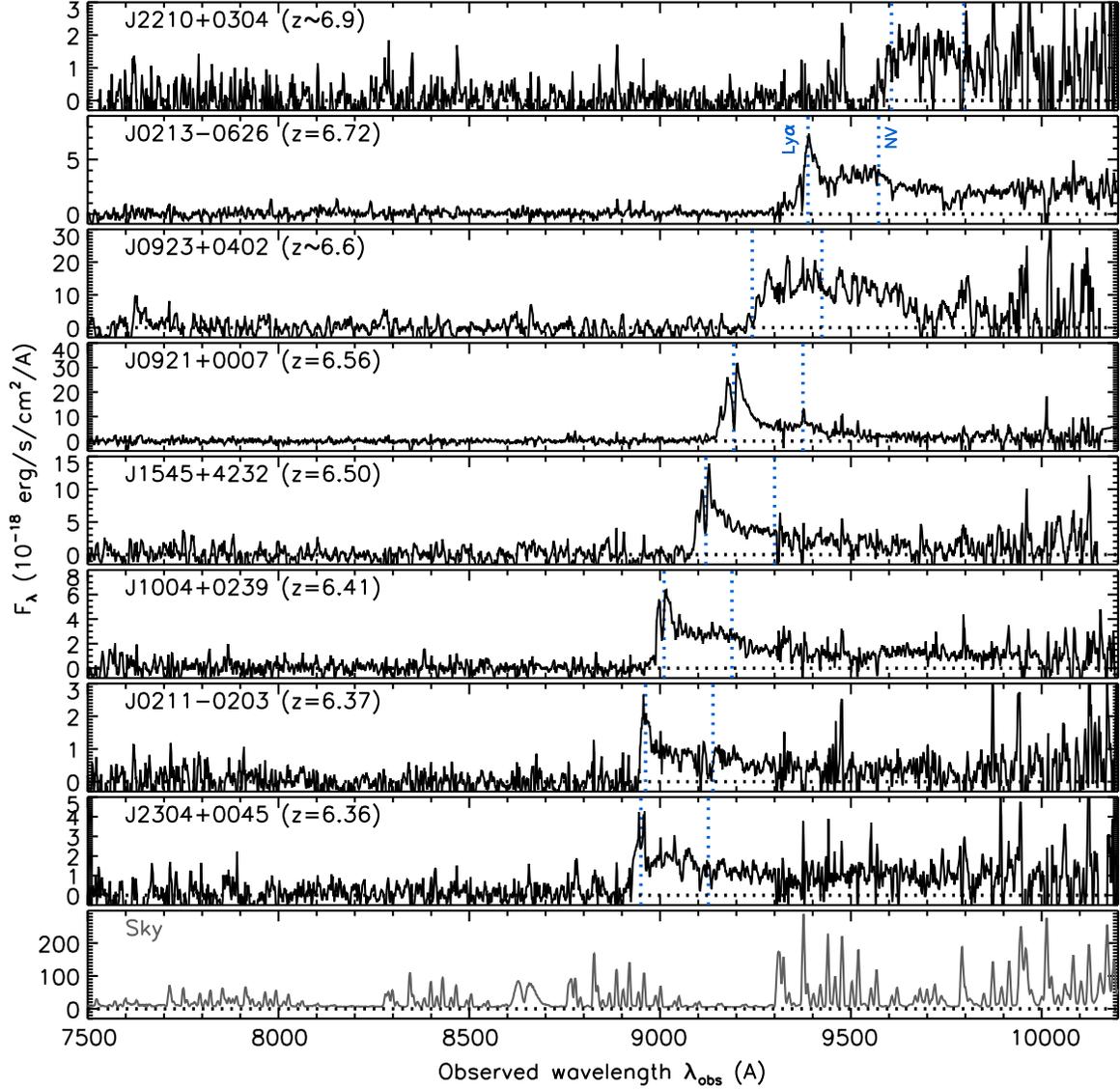}
\caption{Reduced spectra of the first set of eight quasars discovered in this work, displayed in decreasing order of redshift.
The object name and the estimated redshift are indicated at the top left corner of each panel.
The blue dotted lines mark the expected positions of the Ly$\alpha$ and \ion{N}{5} $\lambda$1240 emission lines, given the redshifts.
The spectra were smoothed using inverse-variance weighted means over 3 -- 11 pixels (depending on the S/N), for display purposes.
The bottom panel displays a sky spectrum, as a guide to the expected noise.
\label{fig:spectra1}}
\end{figure*}

\begin{figure*}
\epsscale{1.0}
\plotone{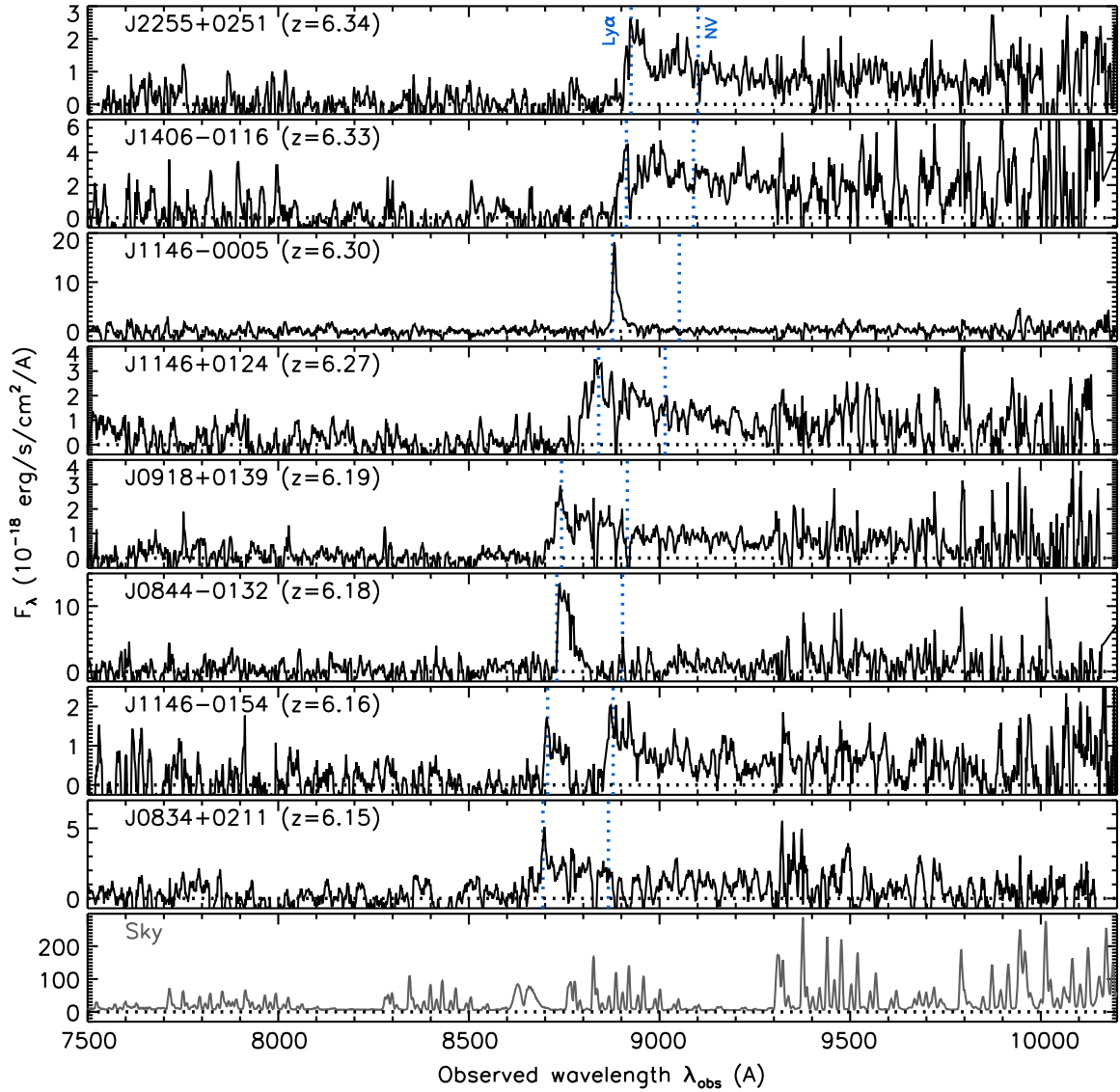}
\caption{Same as Figure \ref{fig:spectra1}, but for the second set of eight quasars.
\label{fig:spectra2}}
\end{figure*}

\begin{figure*}
\epsscale{1.0}
\plotone{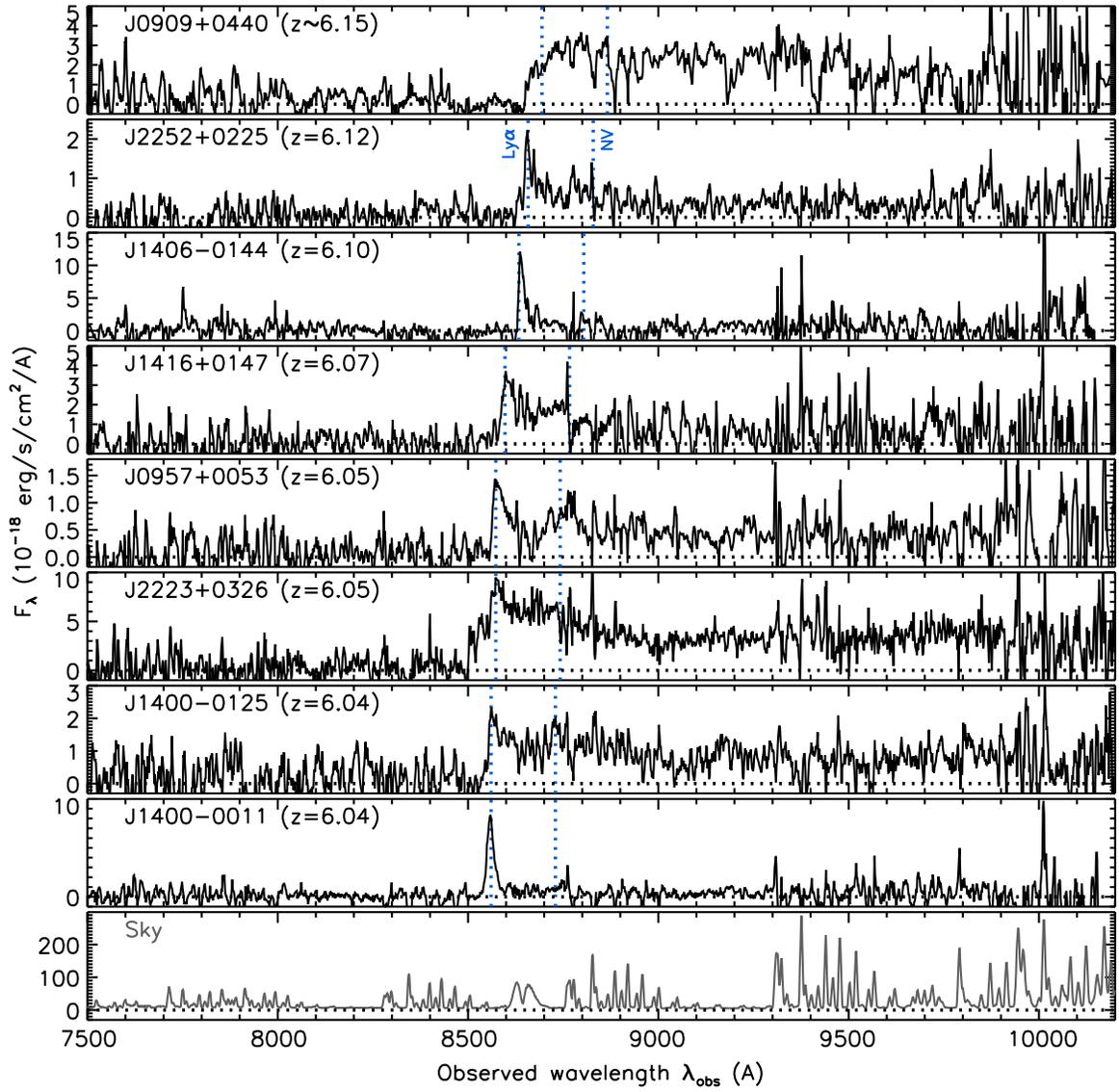}
\caption{Same as Figure \ref{fig:spectra1}, but for the third set of eight quasars.
\label{fig:spectra3}}
\end{figure*}

\begin{figure*}
\epsscale{1.0}
\plotone{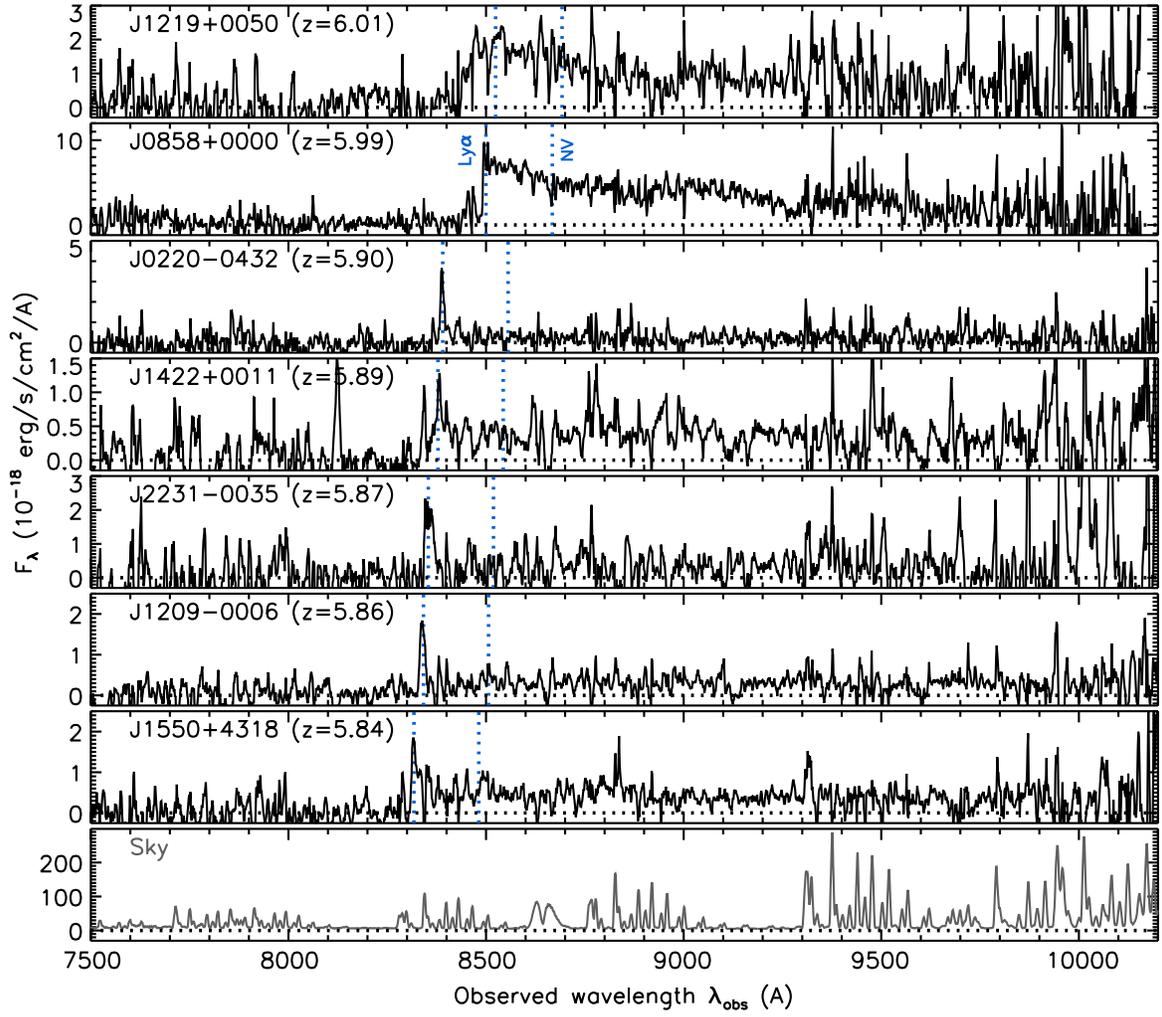}
\caption{Same as Figure \ref{fig:spectra1}, but for the last set of seven quasars.
\label{fig:spectra4}}
\end{figure*}

\begin{figure*}
\epsscale{1.0}
\plotone{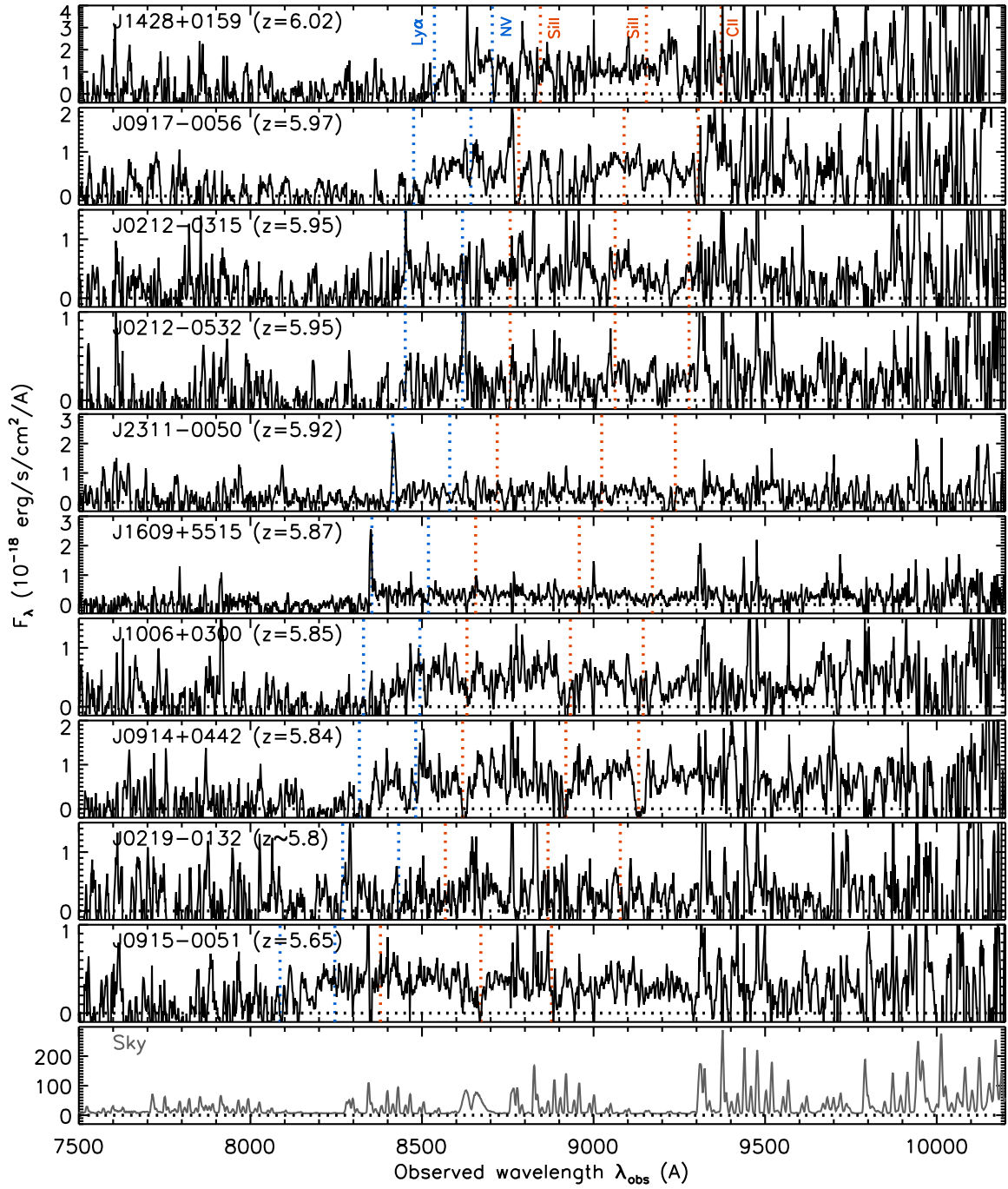}
\caption{Same as Figure \ref{fig:spectra1}, but for the ten high-$z$ galaxies.
The expected positions of the interstellar absorption lines of \ion{Si}{2} $\lambda$1260, \ion{Si}{2} $\lambda$1304, and \ion{C}{2} $\lambda$1335 
are marked by the red dotted lines.
\label{fig:spectra5}}
\end{figure*}

\begin{figure*}
\epsscale{1.0}
\plotone{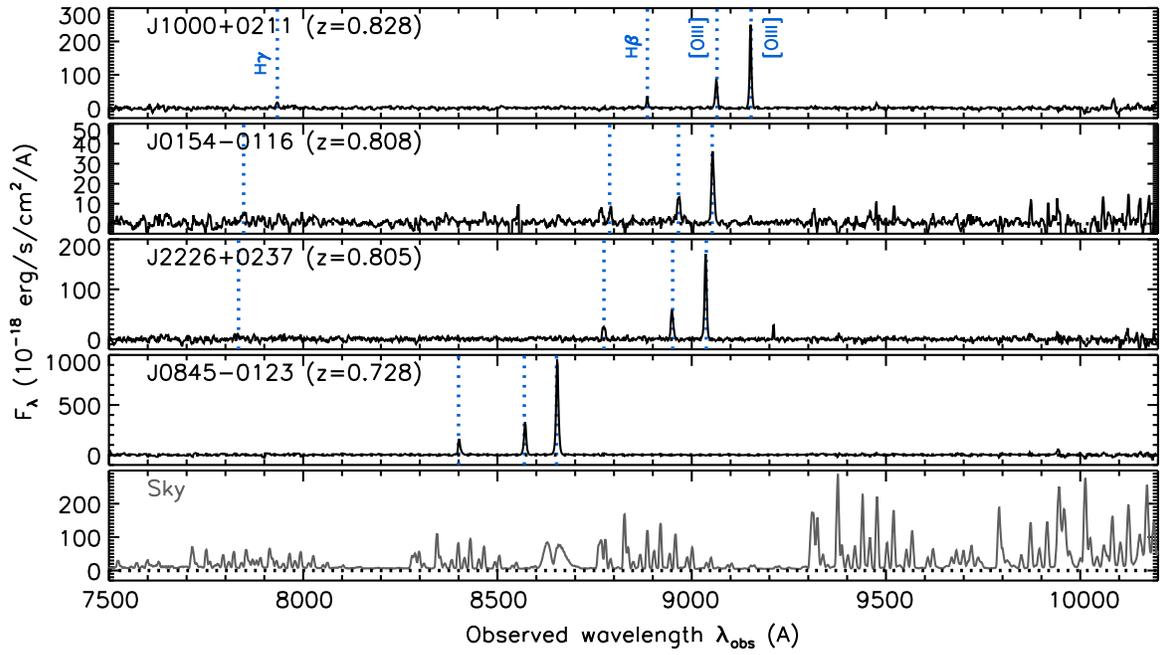}
\caption{Same as Figure \ref{fig:spectra1}, but for the four [\ion{O}{3}] emitters at $z \sim 0.8$.
The expected positions of H$\gamma$, H$\beta$, and two [\ion{O}{3}] lines ($\lambda$4959 and $\lambda$5007) are marked by the dotted lines.
\label{fig:spectra6}}
\end{figure*}

\begin{figure*}
\epsscale{1.0}
\plotone{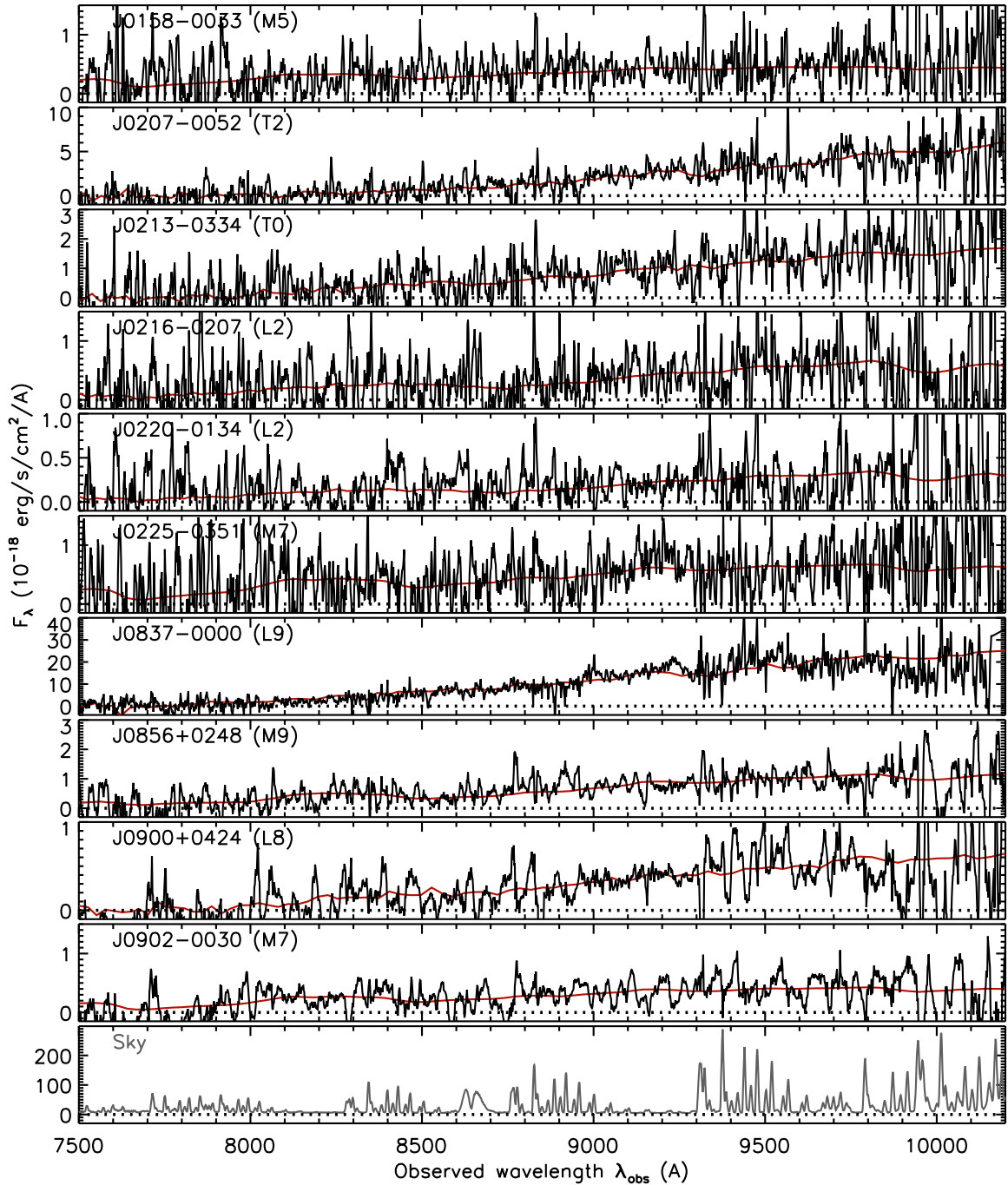}
\caption{Same as Figure \ref{fig:spectra1}, but for the first set of ten cool dwarfs. 
The red lines represent the best-fit templates, whose spectral types are indicated at the top left corner of each panel.
The small-scale ($<$100 \AA) features seen in the spectra are due to noise, given the low S/N.
\label{fig:spectra7}}
\end{figure*}

\begin{figure*}
\epsscale{1.0}
\plotone{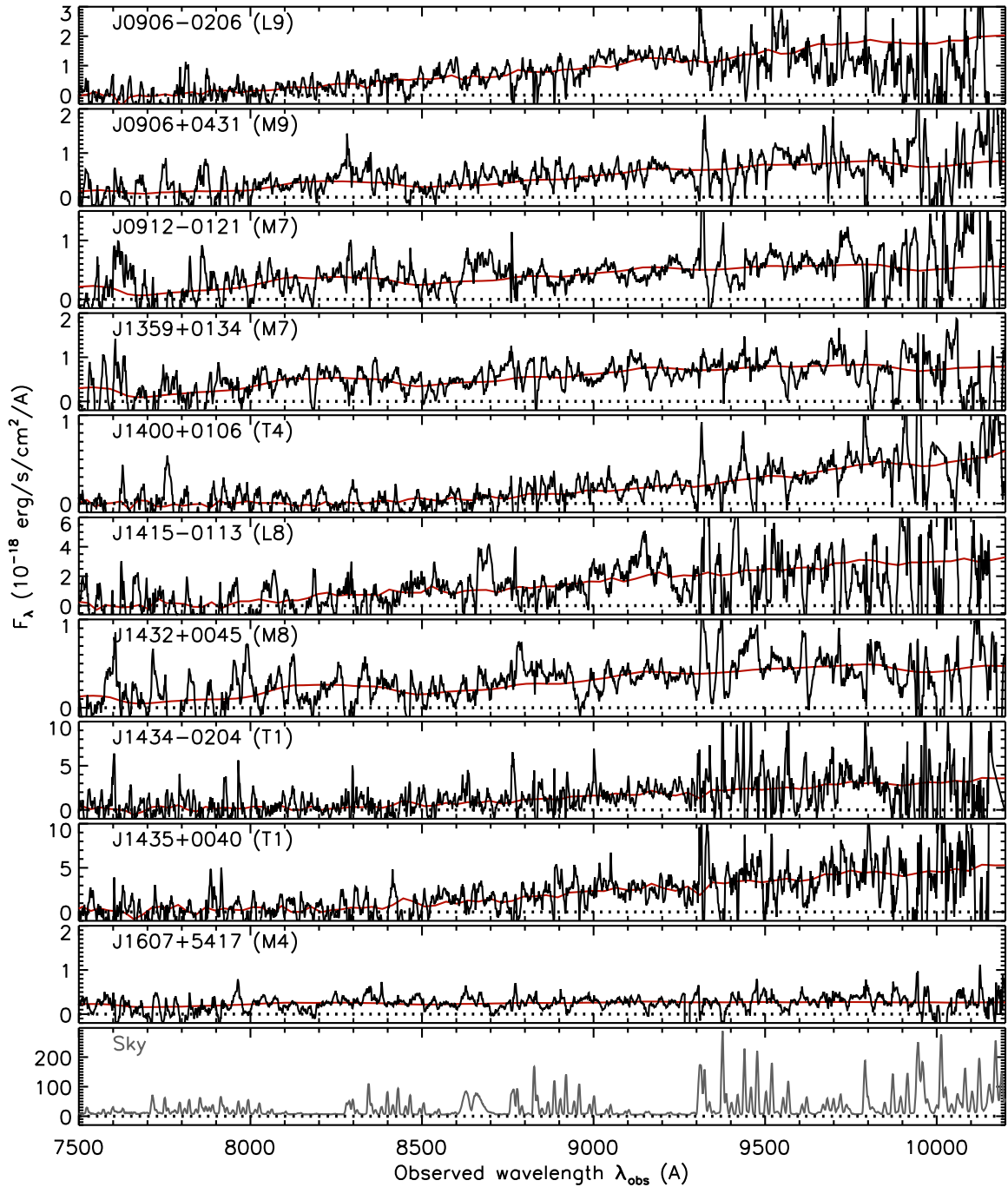}
\caption{Same as Figure \ref{fig:spectra7}, but for the second set of ten cool dwarfs.
\label{fig:spectra8}}
\end{figure*}

\begin{figure*}
\epsscale{1.0}
\plotone{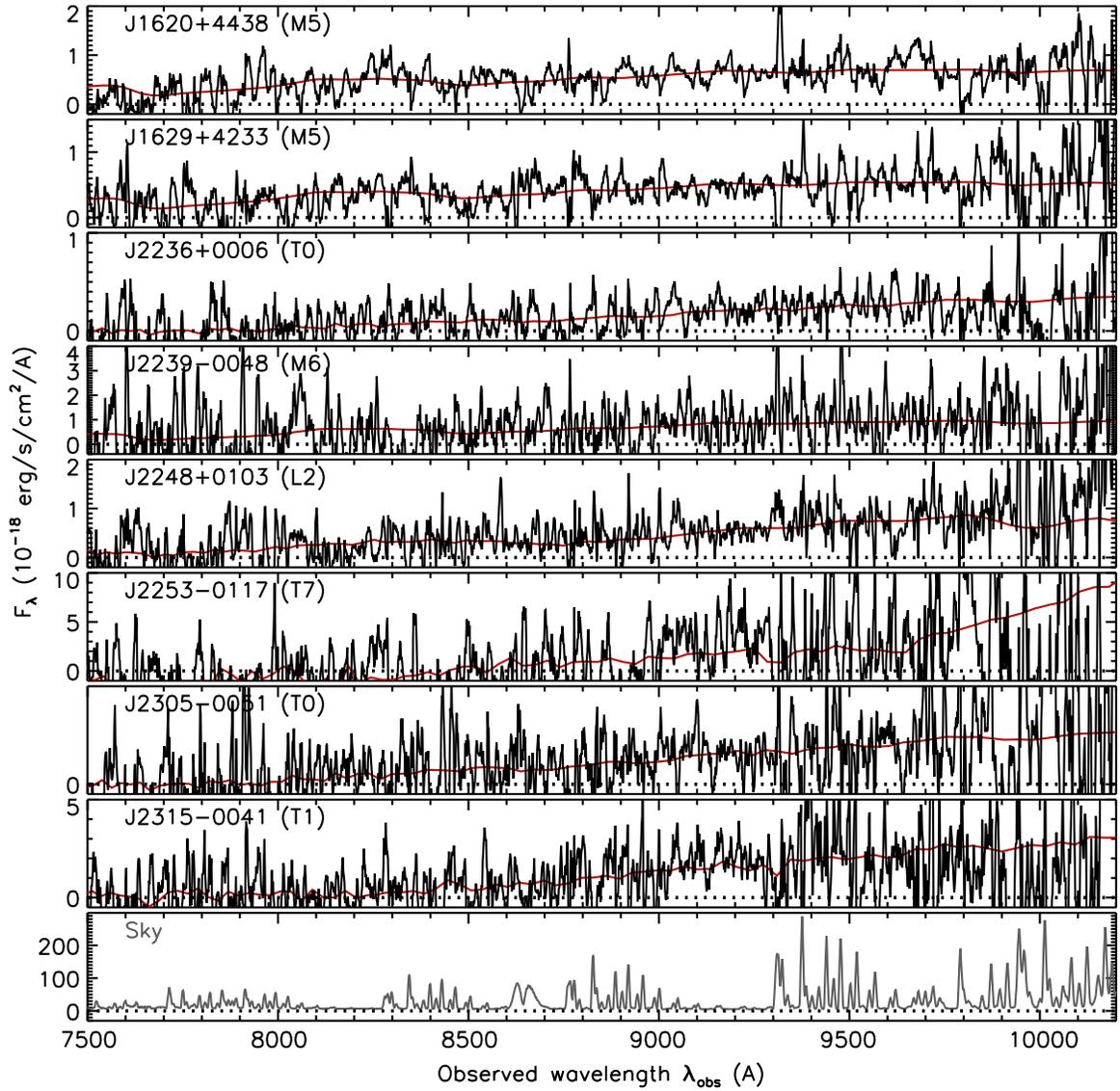}
\caption{Same as Figure \ref{fig:spectra7}, but for the last set of eight cool dwarfs.
\label{fig:spectra9}}
\end{figure*}

Figures \ref{fig:spectra1} -- \ref{fig:spectra9} present the reduced spectra of the 73 candidates we observed.
They include 31 high-$z$ quasars, 10 high-$z$ galaxies, 4 strong [\ion{O}{3}] emitters at $z \sim 0.8$, and 28 cool dwarfs (low-mass stars and brown dwarfs).
Their photometric properties are summarized in Table \ref{tab:photometry}.
The astrometric accuracy of the HSC-SSP data is estimated to be $\lesssim$ 0\arcsec.1 \citep[root mean square;][]{aihara17_pdr1}.
Ten of these objects are detected in the $J$, $H$, and/or $K$ band, as summarized in Table \ref{tab:nir_photometry}.

We identified 31 new quasars at $5.8 < z \le 6.9$, as displayed in Figures \ref{fig:spectra1} -- \ref{fig:spectra4} and listed in the first section of Table \ref{tab:spectroscopy}.
The two highest-$z$ quasars, $J2210+0304$ and $J0213-0626$ at $z = 6.9$ and $z = 6.72$, respectively, are observed as complete $z$-band dropouts on the HSC images.
$J0923+0402$ has a very red $z-y$ color, but is clearly visible in the $z$ band.
All the remaining quasars are $i$-band dropouts. 
It is worth mentioning that we discovered two quasars at $z \sim 6.5$ ($J0921+0007$ and $J1545+4232$), where quasars have similar optical colors to Galactic brown dwarfs and
are thus hard to identify (see Figure 1 of Paper II).
These two quasars have very strong Ly$\alpha$ $+$ \ion{N}{5} $\lambda$1240 lines, which allowed us to separate them from brown dwarfs due to their unusually blue $z-y$ colors.

The majority of the objects in Figures \ref{fig:spectra1} -- \ref{fig:spectra4}
exhibit characteristic spectral features of high-$z$ quasars, namely, strong and 
broad Ly$\alpha$ and in some cases \ion{N}{5} $\lambda$1240, blue rest-UV continua, and sharp continuum breaks just blueward of Ly$\alpha$.
However, for some objects, the quality and wavelength coverage of our spectra are not sufficient to provide robust classification.
In Paper I and II, we reported objects with luminous 
and narrow 
Ly$\alpha$ emission, whose quasar/galaxy classification is still ambiguous.
Our new quasar sample also includes two such objects, namely, $J0220-0432$ and $J1209-0006$.
Following Paper I and II, we classify these objects as possible quasars, due to their high Ly$\alpha$ luminosity ($L > 10^{43}$ erg s$^{-1}$). 
This is based on the fact that, at $z \sim 2$, the majority of Ly$\alpha$ emitters with $L > 10^{43}$ erg s$^{-1}$ are associated with AGNs, based on their X-ray, UV, and radio properties
\citep{konno16}.

We estimated the redshifts of the discovered quasars from the Ly$\alpha$ lines, assuming that the observed line peaks correspond to the intrinsic Ly$\alpha$ wavelength 
(1216 \AA\ in the rest frame).
This assumption is not always correct, due to the strong \ion{H}{1} absorption from the IGM.
It is more difficult to measure redshifts of quasars without clear Ly$\alpha$ lines, as is the case for $J2210+0304$, $J0923+0402$, and several other quasars in our sample.
In these cases we estimated rough redshifts from the wavelengths of the onset of the Gunn-Peterson trough.
Therefore, the redshifts presented here (Table \ref{tab:spectroscopy}) are accompanied by relatively large uncertainties (up to ${\Delta}z \sim 0.1$), and must be interpreted with caution. 
Future follow-up observations at other wavelengths, e.g., near-IR covering \ion{Mg}{2} $\lambda$2800 or sub-mm covering [\ion{C}{2}] 158$\mu$m, are crucial for secure redshift measurements.

The absolute magnitudes $M_{1450}$ and Ly$\alpha$ line properties were measured in the same way as described in Paper II.
For every object, we defined a continuum window at wavelengths relatively free from strong sky emission lines, and extrapolated the measured continuum flux to estimate $M_{1450}$. 
A power-law continuum model with a slope $\alpha = -1.5$ ($F_{\lambda} \propto \lambda^{\alpha}$; e.g., \cite{vandenberk01}) was assumed.
Since the continuum windows (falling in the range of $\lambda_{\rm rest}$ = 1250 -- 1350 \AA) are close to $\lambda_{\rm rest}$ = 1450 \AA, these measurements are not sensitive to the exact value of $\alpha$.
Ly$\alpha$ properties (luminosity, FWHM, and rest-frame equivalent width [EW]) of an object with weak continuum emission, such as $J1146-0005$, was measured with a local continuum defined as an average flux of all the pixels redward of Ly$\alpha$.
For the remaining objects with relatively strong continuum, we measured the properties of the broad Ly$\alpha$ + \ion{N}{5} complex, with a local continuum defined by the above power-law model.
We didn't assume any line profile models, but used the continuum-subtracted flux counts directly to measure these line properties. 
Due to the difficulty in defining accurate continuum levels, these line measurements should be regarded as only approximate.
The resultant line properties are summarized in Table \ref{tab:spectroscopy}.
More detailed descriptions of the above procedure may be found in Paper II.

The ten objects without a broad or luminous ($> 10^{43}$ erg s$^{-1}$) Ly$\alpha$ line, as presented in Figure \ref{fig:spectra5}, 
are most likely galaxies at $z \sim 6$.
We have now spectroscopically identified 24 such high-$z$ luminous galaxies, when combined with the similar objects presented in the previous papers.
Redshifts of these objects were estimated from the observed positions of Ly$\alpha$ emission, the interstellar absorption lines of \ion{Si}{2} $\lambda$1260, 
\ion{Si}{2} $\lambda$1304, \ion{C}{2} $\lambda$1335, and/or the continuum break caused by the IGM \ion{H}{1} absorption.
This is not always easy with our limited S/N, and hence  
the redshifts reported here must be regarded as only approximate (with uncertainties up to ${\Delta}z \sim 0.1$). 
The absolute magnitude  $M_{1450}$ and Ly$\alpha$ properties were measured in the same way as for quasars, except that we assumed a continuum slope of
$\beta = -2.0$ \citep[$F_{\lambda} \propto \lambda^{\beta}$;][]{stanway05}.


This paper also adds four [\ion{O}{3}] emitters at $z \sim 0.8$ (see Figures \ref{fig:spectra6}) to the two similar objects reported in Paper II.
We measured the line properties of H$\gamma$, H$\beta$, [\ion{O}{3}] $\lambda$4959 and $\lambda$5007, and listed the results in Table \ref{tab:spectroscopy}.
Since they have very weak continua, we estimated the continuum levels by summing up all the available pixels except for the above emission lines. 
This still gives relatively large uncertainties in the EWs for some of the objects.
As we discussed in Paper II, their very high [\ion{O}{3}] $\lambda$5007/H$\beta$ ratios may indicate that these are galaxies with sub-solar metallicity and 
high ionization state of the interstellar medium \citep[e.g.,][]{kewley16}, and/or AGN contribution.
$J1000+0211$ and $J0845-0123$ have extremely large [\ion{O}{3}] $\lambda$5007 EWs ($\ga 5000$ \AA), which even exceeds those found in the so-called ``green pea" galaxies
characterized by strong  [\ion{O}{3}] $\lambda$5007 \citep[EW $\la$ 1000 \AA;][]{cardamone09}.
These objects are clearly an interesting subject of future follow-up studies.

Finally, we found 28 new Galactic cool dwarfs (low-mass stars and brown dwarfs), as presented in Figures \ref{fig:spectra7} -- \ref{fig:spectra9}.
We estimated their rough spectral classes by fitting the spectral standard templates of M4- to T8-type dwarfs, taken from the SpeX Prism Spectral Library \citep{burgasser14, skrzypek15}, 
to the observed spectra at $\lambda = 7500 - 9800$ \AA.
The results are summarized in Table \ref{tab:bdtypes} and plotted in the figures.
Due to the low S/N and limited wavelength coverage of the spectra, the classification presented here is accompanied by large uncertainties, and should be regarded as only approximate.

\begin{figure*}
\epsscale{1.0}
\plotone{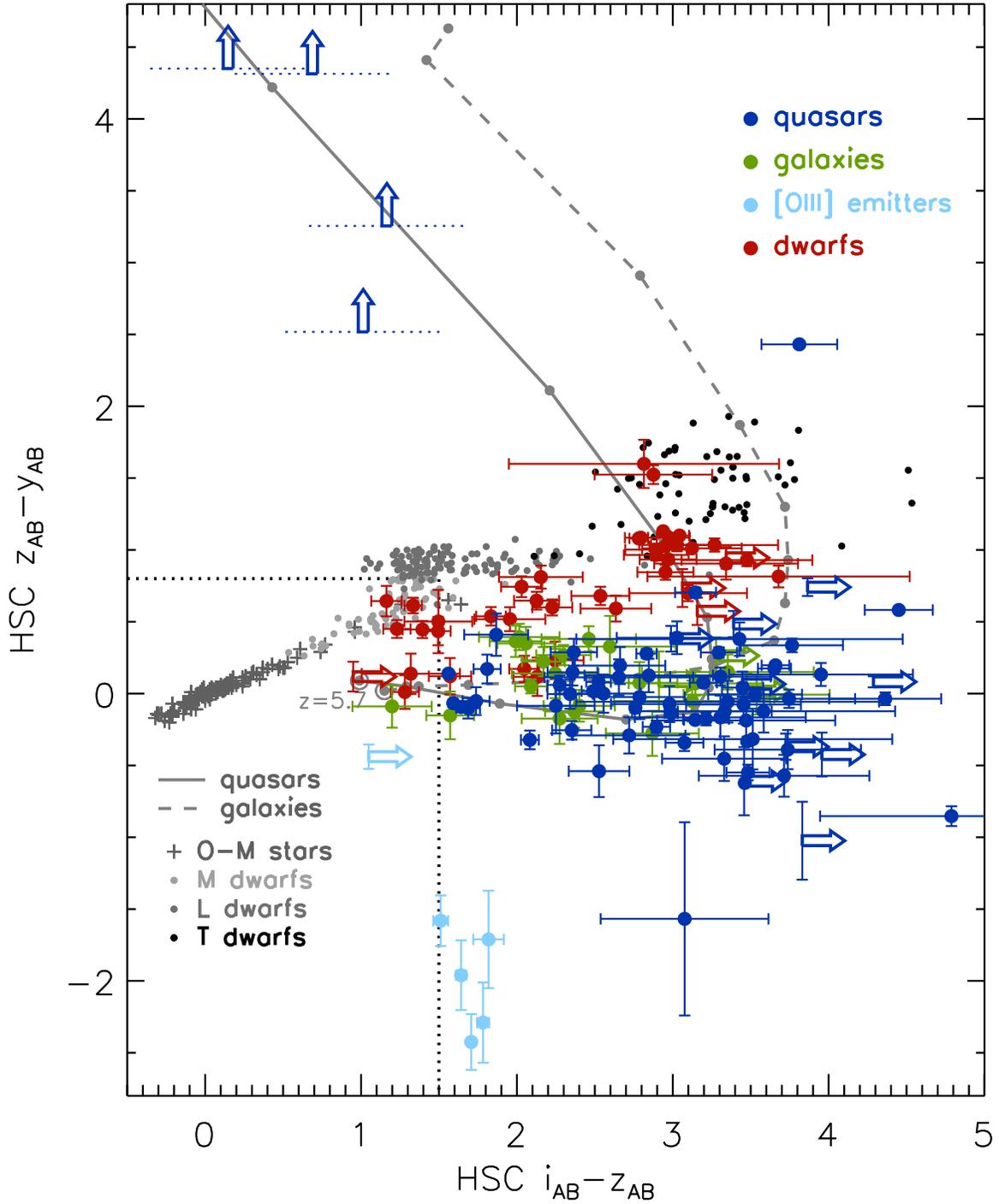}
\caption{
HSC $i_{\rm AB} - z_{\rm AB}$ and $z_{\rm AB} - y_{\rm AB}$ colors of the SHELLQs quasars (blue dots; including previously-known quasars we recovered), 
galaxies (green dots), [\ion{O}{3}] emitters (light blue dots), and cool dwarfs (red dots). 
The grey crosses and dots represent Galactic stars \citep{pickles98} and brown dwarfs, while the solid and dashed lines represent models of quasars and galaxies at 
$z \ge 5.7$; the dots along the lines represent redshifts in steps of 0.1, with $z = 5.7$ marked by the large open circles.
The dotted lines represent the color criteria used in our HSC database query. 
The four quasars at the top-left corner (marked by the up arrows) are undetected in the $i$ and $z$ bands, 
and are plotted at arbitrary horizontal positions.
All the sources discovered in Paper I, II and this work are included.
\label{fig:colordiagram_observed}}
\end{figure*}

In Figure  \ref{fig:colordiagram_observed}, we plot the HSC $i_{\rm AB} - z_{\rm AB}$ and $z_{\rm AB} - y_{\rm AB}$ colors of all the 
spectroscopically-identified objects in Paper I, II, and this work.
We also include the ten previously-known quasars recovered by SHELLQs.
The figure demonstrates that the discovered quasars broadly follow the expected colors from our SED model, while there are outliers with 
very blue $z_{\rm AB} - y_{\rm AB}$ colors, due to exceptionally large Ly$\alpha$ EWs. 
On the other hand, we are finding many cool dwarfs that are bluer than our fiducial models, due to the nature of our selection criteria.
It may be worth noting a clump of brown dwarfs at ($i_{\rm AB} - z_{\rm AB}$, $z_{\rm AB} - y_{\rm AB}$) $\sim$ (3, 1); they lie exactly
on the quasar SED track in the color space, and are thus assigned with high $P_{\rm Q}^{\rm B}$ values as candidates of quasars at $z \sim 6.5$.
The $z \sim 6$ galaxies occupy colors in between quasars and cool dwarfs, due to their weaker Ly$\alpha$ emission than quasars and bluer
continua than cool dwarfs.
Finally, the six [\ion{O}{3}] emitters are well separated from the quasars on this diagram, and do not satisfy our latest selection condition of $i_{\rm AB} - z_{\rm AB} > 2.0$.

\begin{figure*}
\epsscale{1.0}
\plotone{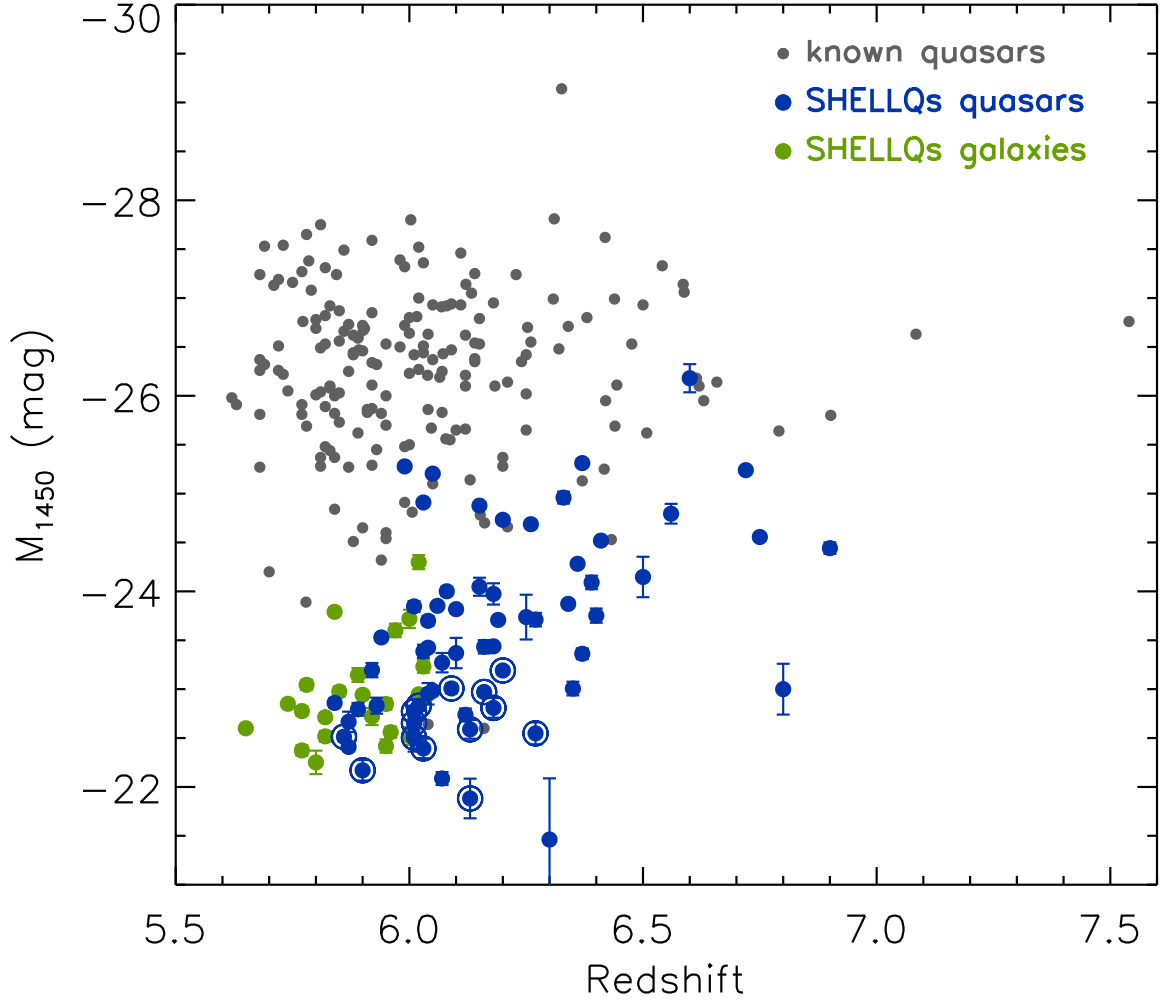}
\caption{Rest-UV absolute magnitude at 1450 \AA\ ($M_{1450}$), as a function of redshift,
	of the SHELLQs quasars (blue dots) and galaxies (green dots), as well as of all the high-$z$ quasars discovered by other surveys and published to date (small grey dots).
	The SHELLQs quasars with narrow Ly$\alpha$ lines are marked with the larger circles.
	All the high-$z$ objects discovered in Paper I, II, and this work are plotted.
\label{fig:luminosity}}
\end{figure*}

Figure \ref{fig:luminosity} displays the redshifts versus absolute magnitudes $M_{1450}$ of the SHELLQs quasars and galaxies, along with those of 
high-$z$ quasars discovered by other surveys.
We continue to probe a new parameter space at $z > 5.7$ and $M_{1450} > -25$ mag, to which other surveys have 
only limited sensitivities.

\begin{figure*}
\epsscale{1.0}
\plotone{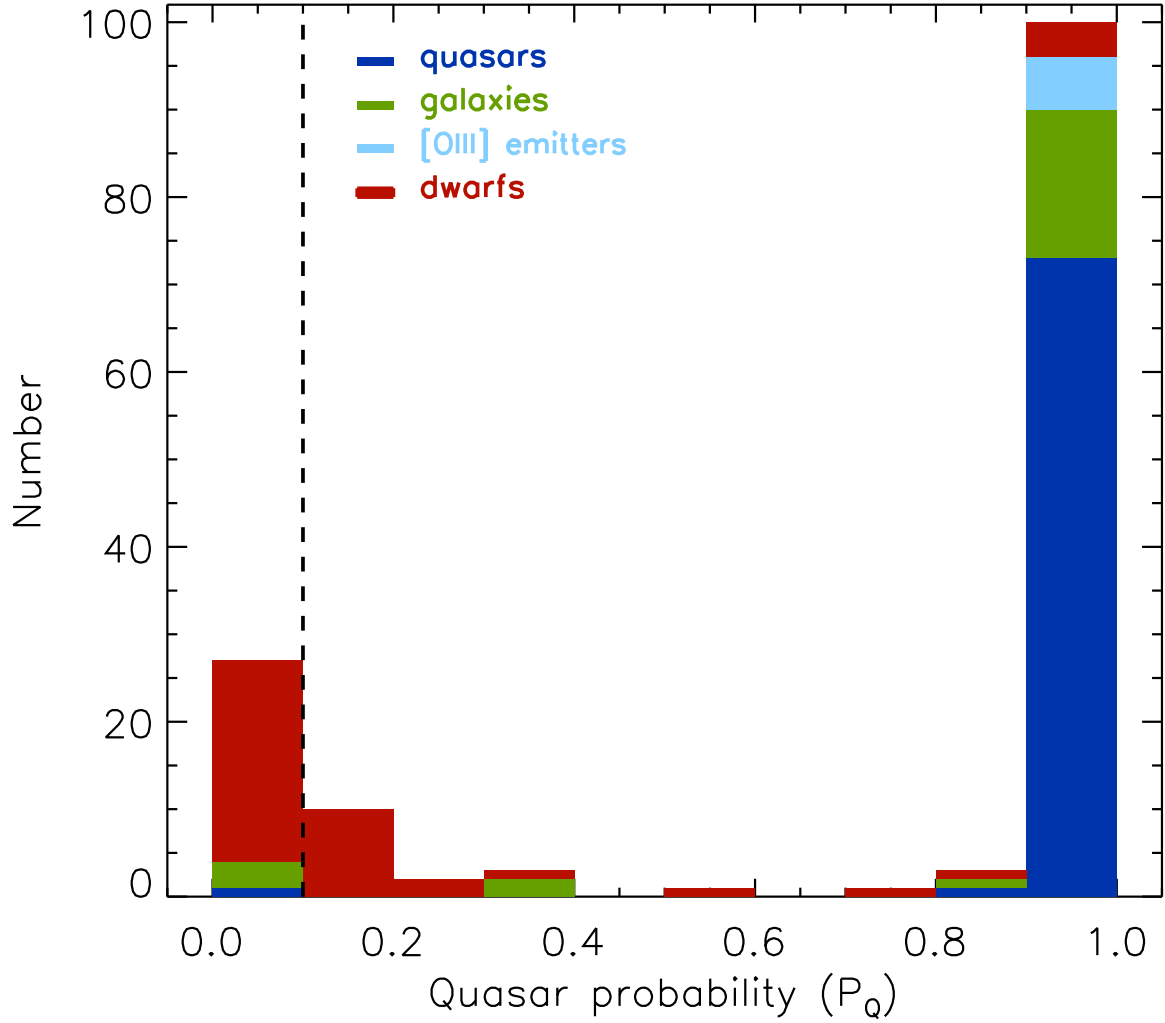}
\caption{Histogram of the Bayesian quasar probability ($P_{\rm Q}^{\rm B}$) of the SHELLQs quasars (blue; including previously-known quasars
we recovered), galaxies (green), [\ion{O}{3}] emitters (light blue), and cool dwarfs (red).
All the sources discovered in Paper I, II, and this work are counted.
The dashed line represents our quasar selection threshold ($P_{\rm Q}^{\rm B} > 0.1$); there are objects below this threshold, 
because they had higher $P_{\rm Q}^{\rm B}$ in the older HSC data releases.
\label{fig:Pq}}
\end{figure*}

Figure \ref{fig:Pq} presents a histogram of the Bayesian quasar probability ($P_{\rm Q}^{\rm B}$), for all the spectroscopically identified objects in Paper I, II, 
and this work.
The $P_{\rm Q}^{\rm B}$ values have a clear bimodal distribution, with the higher peak at $P_{\rm Q}^{\rm B} \sim 1.0$ being dominated by high-$z$ quasars. 
All but one quasar have $P_{\rm Q}^{\rm B} > 0.8$, which implies that their spectral diversity is reasonably covered by the quasar model in our Bayesian algorithm.
The lower peak at $P_{\rm Q}^{\rm B} \sim 0.0$ is populated mostly by cool dwarfs.
Many of these dwarfs lie below the quasar selection threshold ($P_{\rm Q}^{\rm B} = 0.1$), due to the improvement of HSC photometry 
with updates of the data reduction pipeline; they were selected for spectroscopic identification
because they had $P_{\rm Q}^{\rm B} > 0.1$ in the older data releases.
This is also the case for the quasar $J0220-0432$, which had $z_{\rm AB} - y_{\rm AB} \sim 0.0$ and $P_{\rm Q}^{\rm B} = 1.0$ previously but happens to 
have $z_{\rm AB} - y_{\rm AB} \sim 0.4$ and $P_{\rm Q}^{\rm B} = 0.0$ in the latest data release.
Because no additional $z$- or $y$-band data were taken for this object since the previous data release, and because this object is located within a few arcsec of a brighter galaxy, 
this discrepancy in color measurements may be due to different deblender treatment in the different versions of the HSC pipeline.


\clearpage
\startlongtable
\begin{deluxetable*}{ccccccc}
\tablecaption{Photometric Properties \label{tab:photometry}}
\tablehead{
\colhead{Name} & \colhead{R.A.} & 
\colhead{Decl.} & \colhead{$i_{\rm AB}$ (mag)} & 
\colhead{$z_{\rm AB}$ (mag)} & \colhead{$y_{\rm AB}$ (mag)} & 
\colhead{$P_{\rm Q}^{\rm B}$}
} 
\startdata
\multicolumn{7}{c}{Quasars}\\\hline
  $J2210+0304$ & 22$^{\rm h}$10$^{\rm m}$27$^{\rm s}$.24 & $+$03$^\circ$04\arcmin28\arcsec.5 & 27.36 $\pm$ 0.78  &          $>$25.03      & 22.94 $\pm$ 0.06 &         1.000    \\ 
  $J0213-0626$ & 02$^{\rm h}$13$^{\rm m}$16$^{\rm s}$.94 & $-$06$^\circ$26\arcmin15\arcsec.2   & 26.19 $\pm$ 0.30  &          $>$25.05      & 21.69 $\pm$ 0.03 &         1.000     \\ 
  $J0923+0402$ & 09$^{\rm h}$23$^{\rm m}$47$^{\rm s}$.12 & $+$04$^\circ$02\arcmin54\arcsec.4 & 26.45 $\pm$ 0.24  & 22.64 $\pm$ 0.02 & 20.21 $\pm$ 0.01 &         1.000    \\ 
  $J0921+0007$ & 09$^{\rm h}$21$^{\rm m}$20$^{\rm s}$.56 & $+$00$^\circ$07\arcmin22\arcsec.9 & 26.28 $\pm$ 0.22  & 21.83 $\pm$ 0.01 & 21.24 $\pm$ 0.01 &         1.000    \\ 
  $J1545+4232$ & 15$^{\rm h}$45$^{\rm m}$05$^{\rm s}$.62 & $+$42$^\circ$32\arcmin11\arcsec.6 & 25.74 $\pm$ 0.12  & 22.45 $\pm$ 0.01 & 22.16 $\pm$ 0.04 &         1.000     \\ 
  $J1004+0239$ & 10$^{\rm h}$04$^{\rm m}$01$^{\rm s}$.36 & $+$02$^\circ$39\arcmin30\arcsec.7 & 26.41 $\pm$ 0.11  & 22.76 $\pm$ 0.01 & 22.58 $\pm$ 0.01 &         1.000      \\ 
  $J0211-0203$ & 02$^{\rm h}$11$^{\rm m}$44$^{\rm s}$.53 & $-$02$^\circ$03\arcmin03\arcsec.9    & 27.47 $\pm$ 1.05 & 24.04 $\pm$ 0.06 & 23.66 $\pm$ 0.10 &         0.999     \\ 
  $J2304+0045$ & 23$^{\rm h}$04$^{\rm m}$22$^{\rm s}$.97 & $+$00$^\circ$45\arcmin05\arcsec.4 & 26.85 $\pm$ 0.33 & 23.08 $\pm$ 0.02 & 22.75 $\pm$ 0.04 &         1.000     \\ 
  $J2255+0251$ & 22$^{\rm h}$55$^{\rm m}$38$^{\rm s}$.04 & $+$02$^\circ$51\arcmin26\arcsec.6 &          $>$25.72      & 23.44 $\pm$ 0.05 & 22.95 $\pm$ 0.07 &         1.000     \\ 
  $J1406-0116$ & 14$^{\rm h}$06$^{\rm m}$29$^{\rm s}$.12 & $-$01$^\circ$16\arcmin11\arcsec.2   & 25.78 $\pm$ 0.13 & 22.64 $\pm$ 0.02 & 21.93 $\pm$ 0.03 &         1.000     \\ 
  $J1146-0005$ & 11$^{\rm h}$46$^{\rm m}$58$^{\rm s}$.89 & $-$00$^\circ$05\arcmin37\arcsec.7   &          $>$26.43      & 23.76 $\pm$ 0.04 & 24.78 $\pm$ 0.27 &         1.000     \\ 
  $J1146+0124$ & 11$^{\rm h}$46$^{\rm m}$48$^{\rm s}$.42 & $+$01$^\circ$24\arcmin20\arcsec.1 & 26.76 $\pm$ 0.42 & 23.01 $\pm$ 0.03 & 23.05 $\pm$ 0.06 &         1.000     \\ 
  $J0918+0139$ & 09$^{\rm h}$18$^{\rm m}$33$^{\rm s}$.17 & $+$01$^\circ$39\arcmin23\arcsec.3 & 26.70 $\pm$ 0.30 & 23.17 $\pm$ 0.03 & 23.18 $\pm$ 0.04 &         1.000     \\ 
  $J0844-0132$ & 08$^{\rm h}$44$^{\rm m}$08$^{\rm s}$.61 & $-$01$^\circ$32\arcmin16\arcsec.5   &          $>$26.24    & 23.31 $\pm$ 0.03 & 23.73 $\pm$ 0.15 &         1.000     \\ 
  $J1146-0154$ & 11$^{\rm h}$46$^{\rm m}$32$^{\rm s}$.66 & $-$01$^\circ$54\arcmin38\arcsec.3   & 26.90 $\pm$ 0.55 & 23.60 $\pm$ 0.06 & 23.76 $\pm$ 0.14 &         1.000     \\ 
  $J0834+0211$ & 08$^{\rm h}$34$^{\rm m}$00$^{\rm s}$.88 & $+$02$^\circ$11\arcmin46\arcsec.9 & 26.27 $\pm$ 0.32 & 22.97 $\pm$ 0.03 & 22.85 $\pm$ 0.05 &         1.000     \\ 
  $J0909+0440$ & 09$^{\rm h}$09$^{\rm m}$21$^{\rm s}$.50 & $+$04$^\circ$40\arcmin42\arcsec.9 & 25.02 $\pm$ 0.06 & 22.18 $\pm$ 0.02 & 21.91 $\pm$ 0.03 &         1.000     \\ 
  $J2252+0225$ & 22$^{\rm h}$52$^{\rm m}$05$^{\rm s}$.44 & $+$02$^\circ$25\arcmin31\arcsec.9 & 26.78 $\pm$ 0.37 & 23.80 $\pm$ 0.06 & 23.88 $\pm$ 0.11 &         1.000     \\ 
  $J1406-0144$ & 14$^{\rm h}$06$^{\rm m}$46$^{\rm s}$.88 & $-$01$^\circ$44\arcmin02\arcsec.6   & 26.78 $\pm$ 0.37 & 23.33 $\pm$ 0.04 & 23.40 $\pm$ 0.13 &         1.000     \\ 
  $J1416+0147$ & 14$^{\rm h}$16$^{\rm m}$53$^{\rm s}$.01 & $+$01$^\circ$47\arcmin02\arcsec.2 & 26.60 $\pm$ 0.55 & 22.89 $\pm$ 0.04 & 23.46 $\pm$ 0.14 &         1.000      \\ 
  $J0957+0053$ & 09$^{\rm h}$57$^{\rm m}$40$^{\rm s}$.39 & $+$00$^\circ$53\arcmin33\arcsec.6 &          $>$26.91     & 23.66 $\pm$ 0.01 & 23.58 $\pm$ 0.04 &         1.000      \\ 
  $J2223+0326$ & 22$^{\rm h}$23$^{\rm m}$09$^{\rm s}$.51 & $+$03$^\circ$26\arcmin20\arcsec.3 & 24.07 $\pm$ 0.03 & 21.31 $\pm$ 0.01 & 21.41 $\pm$ 0.02 &         1.000     \\ 
  $J1400-0125$ & 14$^{\rm h}$00$^{\rm m}$29$^{\rm s}$.99 & $-$01$^\circ$25\arcmin21\arcsec.0   & 25.16 $\pm$ 0.08 & 22.88 $\pm$ 0.02 & 22.82 $\pm$ 0.06 &         1.000     \\ 
  $J1400-0011$ & 14$^{\rm h}$00$^{\rm m}$28$^{\rm s}$.79 & $-$00$^\circ$11\arcmin51\arcsec.5    & 26.15 $\pm$ 0.19 & 23.62 $\pm$ 0.04 & 24.16 $\pm$ 0.18 &         1.000     \\ 
  $J1219+0050$ & 12$^{\rm h}$19$^{\rm m}$05$^{\rm s}$.34 & $+$00$^\circ$50\arcmin37\arcsec.5 & 24.90 $\pm$ 0.05 & 22.81 $\pm$ 0.02 & 23.13 $\pm$ 0.06 &         1.000     \\ 
  $J0858+0000$ & 08$^{\rm h}$58$^{\rm m}$13$^{\rm s}$.51 & $+$00$^\circ$00\arcmin57\arcsec.1 & 24.28 $\pm$ 0.04 & 21.29 $\pm$ 0.01 & 21.43 $\pm$ 0.01 &         1.000     \\ 
  $J0220-0432$ & 02$^{\rm h}$20$^{\rm m}$29$^{\rm s}$.72 & $-$04$^\circ$32\arcmin04\arcsec.0   & 26.29 $\pm$ 0.18 & 24.42 $\pm$ 0.10 & 24.01 $\pm$ 0.11 &         0.000    \\ 
  $J1422+0011$ & 14$^{\rm h}$22$^{\rm m}$00$^{\rm s}$.23 & $+$00$^\circ$11\arcmin03\arcsec.0 & 26.22 $\pm$ 0.14  & 23.85 $\pm$ 0.05 & 23.56 $\pm$ 0.08 &         1.000     \\ 
  $J2231-0035$ & 22$^{\rm h}$31$^{\rm m}$48$^{\rm s}$.89 & $-$00$^\circ$35\arcmin47\arcsec.5   & 26.52 $\pm$ 0.24 & 24.27 $\pm$ 0.07 & 24.35 $\pm$ 0.18 &         1.000    \\ 
  $J1209-0006$ & 12$^{\rm h}$09$^{\rm m}$23$^{\rm s}$.99 & $-$00$^\circ$06\arcmin46\arcsec.5   & 26.69 $\pm$ 0.23 & 24.13 $\pm$ 0.05 & 24.13 $\pm$ 0.12 &         1.000     \\ 
  $J1550+4318$ & 15$^{\rm h}$50$^{\rm m}$00$^{\rm s}$.93 & $+$43$^\circ$18\arcmin02\arcsec.8 & 25.52 $\pm$ 0.08 & 23.71 $\pm$ 0.03 & 23.54 $\pm$ 0.10 &         1.000     \\ 
 \hline\multicolumn{7}{c}{Galaxies}\\\hline
  $J1428+0159$ & 14$^{\rm h}$28$^{\rm m}$24$^{\rm s}$.71 & $+$01$^\circ$59\arcmin34\arcsec.4 & 26.05 $\pm$ 0.43 & 22.90 $\pm$ 0.05 & 22.83 $\pm$ 0.07 &         1.000      \\ 
  $J0917-0056$ & 09$^{\rm h}$17$^{\rm m}$00$^{\rm s}$.28 & $-$00$^\circ$56\arcmin58\arcsec.1   & 26.06 $\pm$ 0.18 & 23.69 $\pm$ 0.05 & 23.81 $\pm$ 0.10 &         1.000     \\ 
  $J0212-0315$ & 02$^{\rm h}$12$^{\rm m}$36$^{\rm s}$.67 & $-$03$^\circ$15\arcmin17\arcsec.6   & 26.98 $\pm$ 0.50 & 23.85 $\pm$ 0.05 & 23.90 $\pm$ 0.13 &         1.000     \\ 
  $J0212-0532$ & 02$^{\rm h}$12$^{\rm m}$49$^{\rm s}$.46 & $-$05$^\circ$32\arcmin38\arcsec.0   &          $>$26.22     & 24.40 $\pm$ 0.08 & 24.31 $\pm$ 0.19 &         1.000    \\ 
  $J2311-0050$ & 23$^{\rm h}$11$^{\rm m}$42$^{\rm s}$.86 & $-$00$^\circ$50\arcmin20\arcsec.7   & 26.83 $\pm$ 0.41 & 24.23 $\pm$ 0.09 & 23.91 $\pm$ 0.20 &         0.356    \\ 
  $J1609+5515$ & 16$^{\rm h}$09$^{\rm m}$52$^{\rm s}$.79 & $+$55$^\circ$15\arcmin48\arcsec.5 & 25.89 $\pm$ 0.08 & 24.15 $\pm$ 0.03 & 24.20 $\pm$ 0.09 &         1.000      \\ 
  $J1006+0300$ & 10$^{\rm h}$06$^{\rm m}$33$^{\rm s}$.53 & $+$03$^\circ$00\arcmin05\arcsec.2 & 25.79 $\pm$ 0.10 & 23.70 $\pm$ 0.02 & 23.65 $\pm$ 0.05 &         1.000      \\ 
  $J0914+0442$ & 09$^{\rm h}$14$^{\rm m}$36$^{\rm s}$.47 & $+$04$^\circ$42\arcmin31\arcsec.7 & 25.46 $\pm$ 0.12 & 23.29 $\pm$ 0.06 & 23.07 $\pm$ 0.08 &         0.980     \\ 
  $J0219-0132$ & 02$^{\rm h}$19$^{\rm m}$30$^{\rm s}$.94 & $-$01$^\circ$32\arcmin07\arcsec.2   & 25.47 $\pm$ 0.24 & 24.27 $\pm$ 0.07 & 24.36 $\pm$ 0.13 &         0.820    \\ 
  $J0915-0051$ & 09$^{\rm h}$15$^{\rm m}$45$^{\rm s}$.02 & $-$00$^\circ$51\arcmin36\arcsec.0   & 25.61 $\pm$ 0.13 & 24.04 $\pm$ 0.08 & 24.19 $\pm$ 0.15 &         0.316     \\ 
 \hline\multicolumn{7}{c}{[\ion{O}{3}] Emitters}\\\hline
  $J1000+0211$ & 10$^{\rm h}$00$^{\rm m}$12$^{\rm s}$.46 & $+$02$^\circ$11\arcmin27\arcsec.4  & 24.65 $\pm$ 0.04 & 22.87 $\pm$ 0.02 & 25.16 $\pm$ 0.28 &         1.000      \\ 
  $J0154-0116$ & 01$^{\rm h}$54$^{\rm m}$31$^{\rm s}$.69 & $-$01$^\circ$16\arcmin19\arcsec.5   &          $>$23.08     & 22.75 $\pm$ 0.04 & 23.19 $\pm$ 0.07 &         1.000     \\ 
  $J2226+0237$ & 22$^{\rm h}$26$^{\rm m}$49$^{\rm s}$.68 & $+$02$^\circ$37\arcmin54\arcsec.0 & 24.25 $\pm$ 0.03 & 22.61 $\pm$ 0.02 & 24.57 $\pm$ 0.24 &         1.000     \\ 
  $J0845-0123$ & 08$^{\rm h}$45$^{\rm m}$16$^{\rm s}$.54 & $-$01$^\circ$23\arcmin21\arcsec.6   & 23.59 $\pm$ 0.02 & 21.89 $\pm$ 0.01 & 24.31 $\pm$ 0.19 &         1.000     \\ 
   \hline\multicolumn{7}{c}{Cool Dwarfs}\\\hline
  $J0158-0033$ & 01$^{\rm h}$58$^{\rm m}$07$^{\rm s}$.10 & $-$00$^\circ$33\arcmin32\arcsec.1   &          $>$23.65     & 23.80 $\pm$ 0.07 & 23.68 $\pm$ 0.07 &         0.119     \\ 
  $J0207-0052$ & 02$^{\rm h}$07$^{\rm m}$01$^{\rm s}$.05 & $-$00$^\circ$52\arcmin25\arcsec.0   & 25.34 $\pm$ 0.32 & 22.22 $\pm$ 0.01 & 21.21 $\pm$ 0.01 &         0.831     \\ 
  $J0213-0334$ & 02$^{\rm h}$13$^{\rm m}$50$^{\rm s}$.82 & $-$03$^\circ$34\arcmin45\arcsec.2  & 25.99 $\pm$ 0.20 & 23.10 $\pm$ 0.03 & 22.10 $\pm$ 0.03 &         0.105     \\ 
  $J0216-0207$ & 02$^{\rm h}$16$^{\rm m}$17$^{\rm s}$.18 & $-$02$^\circ$07\arcmin19\arcsec.3   & 26.14 $\pm$ 0.26 & 23.99 $\pm$ 0.06 & 23.18 $\pm$ 0.06 &         0.000     \\ 
  $J0220-0134$ & 02$^{\rm h}$20$^{\rm m}$47$^{\rm s}$.41 & $-$01$^\circ$34\arcmin50\arcsec.6   & 25.73 $\pm$ 0.31 & 24.41 $\pm$ 0.08 & 24.27 $\pm$ 0.12 &         0.042    \\ 
  $J0225-0351$ & 02$^{\rm h}$25$^{\rm m}$00$^{\rm s}$.18 & $-$03$^\circ$51\arcmin46\arcsec.8   & 25.33 $\pm$ 0.08 & 23.50 $\pm$ 0.05 & 22.96 $\pm$ 0.05 &         0.000    \\ 
  $J0837-0000$ & 08$^{\rm h}$37$^{\rm m}$17$^{\rm s}$.18 & $-$00$^\circ$00\arcmin21\arcsec.0   & 23.18 $\pm$ 0.01 & 20.24 $\pm$ 0.01 & 19.11 $\pm$ 0.01 &         0.000     \\ 
  $J0856+0248$ & 08$^{\rm h}$56$^{\rm m}$09$^{\rm s}$.10 & $+$02$^\circ$48\arcmin51\arcsec.1 & 24.63 $\pm$ 0.12 & 23.24 $\pm$ 0.03 & 22.79 $\pm$ 0.05 &         0.000     \\ 
  $J0900+0424$ & 09$^{\rm h}$00$^{\rm m}$18$^{\rm s}$.40 & $+$04$^\circ$24\arcmin15\arcsec.5 & 27.46 $\pm$ 0.45 & 24.11 $\pm$ 0.08 & 23.21 $\pm$ 0.08 &         0.021     \\ 
  $J0902-0030$ & 09$^{\rm h}$02$^{\rm m}$22$^{\rm s}$.31 & $-$00$^\circ$30\arcmin40\arcsec.4   & 26.02 $\pm$ 0.22 & 24.07 $\pm$ 0.05 & 23.55 $\pm$ 0.07 &         0.000     \\ 
  $J0906-0206$ & 09$^{\rm h}$06$^{\rm m}$50$^{\rm s}$.82 & $-$02$^\circ$06\arcmin10\arcsec.2   & 25.06 $\pm$ 0.14 & 22.93 $\pm$ 0.05 & 22.29 $\pm$ 0.04 &         0.000     \\ 
  $J0906+0431$ & 09$^{\rm h}$06$^{\rm m}$55$^{\rm s}$.10 & $+$04$^\circ$31\arcmin31\arcsec.9 & 25.05 $\pm$ 0.06 & 23.56 $\pm$ 0.06 & 23.12 $\pm$ 0.08 &         0.000     \\ 
  $J0912-0121$ & 09$^{\rm h}$12$^{\rm m}$10$^{\rm s}$.99 & $-$01$^\circ$21\arcmin02\arcsec.9   &          $>$25.77     & 23.74 $\pm$ 0.07 & 23.17 $\pm$ 0.07 &         0.994     \\ 
  $J1359+0134$ & 13$^{\rm h}$59$^{\rm m}$45$^{\rm s}$.00 & $+$01$^\circ$34\arcmin12\arcsec.3 & 24.60 $\pm$ 0.07 & 23.32 $\pm$ 0.06 & 23.31 $\pm$ 0.10 &         0.533     \\ 
  $J1400+0106$ & 14$^{\rm h}$00$^{\rm m}$15$^{\rm s}$.38 & $+$01$^\circ$06\arcmin23\arcsec.7 & 27.91 $\pm$ 0.85 & 25.10 $\pm$ 0.15 & 23.50 $\pm$ 0.07 &         0.000     \\ 
  $J1415-0113$ & 14$^{\rm h}$15$^{\rm m}$32$^{\rm s}$.20 & $-$01$^\circ$13\arcmin14\arcsec.3   & 25.08 $\pm$ 0.06 & 22.30 $\pm$ 0.01 & 21.22 $\pm$ 0.01 &         0.000    \\ 
  $J1432+0045$ & 14$^{\rm h}$32$^{\rm m}$05$^{\rm s}$.78 & $+$00$^\circ$45\arcmin31\arcsec.7 & 25.59 $\pm$ 0.12 & 24.01 $\pm$ 0.07 & 23.89 $\pm$ 0.11 &         0.027     \\ 
  $J1434-0204$ & 14$^{\rm h}$34$^{\rm m}$37$^{\rm s}$.36 & $-$02$^\circ$04\arcmin16\arcsec.7   & 25.33 $\pm$ 0.18 & 22.36 $\pm$ 0.03 & 21.42 $\pm$ 0.02 &         0.000    \\ 
  $J1435+0040$ & 14$^{\rm h}$35$^{\rm m}$29$^{\rm s}$.69 & $+$00$^\circ$40\arcmin35\arcsec.3 & 25.17 $\pm$ 0.06 & 22.12 $\pm$ 0.01 & 21.02 $\pm$ 0.01 &         0.000    \\ 
  $J1607+5417$ & 16$^{\rm h}$07$^{\rm m}$32$^{\rm s}$.84 & $+$54$^\circ$17\arcmin24\arcsec.5 & 26.46 $\pm$ 0.12 & 24.23 $\pm$ 0.03 & 23.63 $\pm$ 0.05 &         0.295      \\ 
  $J1620+4438$ & 16$^{\rm h}$20$^{\rm m}$46$^{\rm s}$.91 & $+$44$^\circ$38\arcmin39\arcsec.7 & 25.64 $\pm$ 0.16 & 23.39 $\pm$ 0.09 & 23.16 $\pm$ 0.10 &         0.154     \\ 
  $J1629+4233$ & 16$^{\rm h}$29$^{\rm m}$48$^{\rm s}$.12 & $+$42$^\circ$33\arcmin38\arcsec.5 & 25.77 $\pm$ 0.09 & 23.72 $\pm$ 0.04 & 23.55 $\pm$ 0.08 &         1.000     \\ 
  $J2236+0006$ & 22$^{\rm h}$36$^{\rm m}$12$^{\rm s}$.42 & $+$00$^\circ$06\arcmin32\arcsec.6 & 26.27 $\pm$ 0.18 & 24.77 $\pm$ 0.13 & 24.27 $\pm$ 0.18 &         0.000    \\ 
  $J2239-0048$ & 22$^{\rm h}$39$^{\rm m}$53$^{\rm s}$.43 & $-$00$^\circ$48\arcmin02\arcsec.6   & 24.23 $\pm$ 0.08 & 23.00 $\pm$ 0.04 & 22.55 $\pm$ 0.05 &         0.000     \\ 
  $J2248+0103$ & 22$^{\rm h}$48$^{\rm m}$34$^{\rm s}$.69 & $+$01$^\circ$03\arcmin12\arcsec.2 & 26.73 $\pm$ 0.38 & 23.63 $\pm$ 0.04 & 22.93 $\pm$ 0.04 &         0.997     \\ 
  $J2253-0117$ & 22$^{\rm h}$53$^{\rm m}$57$^{\rm s}$.80 & $-$01$^\circ$17\arcmin05\arcsec.3   & 25.65 $\pm$ 0.21 & 23.01 $\pm$ 0.08 & 22.42 $\pm$ 0.04 &         0.347     \\ 
  $J2305-0051$ & 23$^{\rm h}$05$^{\rm m}$05$^{\rm s}$.75 & $-$00$^\circ$51\arcmin32\arcsec.6   & 25.71 $\pm$ 0.13 & 23.58 $\pm$ 0.09 & 23.47 $\pm$ 0.10 &         0.719     \\ 
  $J2315-0041$ & 23$^{\rm h}$15$^{\rm m}$14$^{\rm s}$.06 & $-$00$^\circ$41\arcmin01\arcsec.1   & 25.69 $\pm$ 0.13 & 22.72 $\pm$ 0.02 & 21.63 $\pm$ 0.02 &         0.195     \\ 
\enddata
\tablecomments{Coordinates are at J2000.0. We took magnitudes from the latest HSC-SSP data release, and recalculated $P_{\rm Q}^{\rm B}$ for objects selected
with the older data releases (this is why the quasar $J0220-0432$ has $P_{\rm Q}^{\rm B} = 0.000$, which used to be higher in the old data release; see text).
Magnitude upper limits are placed at $5\sigma$ significance.}
\end{deluxetable*}

\begin{deluxetable*}{ccccccc}
\tablecaption{$JHK$ magnitudes of the objects detected in UKIDSS or VIKING \label{tab:nir_photometry}}
\tablehead{
\colhead{} & \multicolumn{3}{c}{UKIDSS} & \multicolumn{3}{c}{VIKING} \\
\colhead{Name} & \colhead{$J_{\rm AB}$} & 
\colhead{$H_{\rm AB}$} & \colhead{$K_{\rm AB}$} & 
\colhead{$J_{\rm AB}$} & 
\colhead{$H_{\rm AB}$} & \colhead{$K_{\rm AB}$}\\
\colhead{} & \colhead{(mag)} & \colhead{(mag)} & \colhead{(mag)} & \colhead{(mag)} & \colhead{(mag)} & \colhead{(mag)} 
} 
\startdata
\multicolumn{7}{c}{Quasars}\\\hline
  $J0923+0402$ & 20.02 $\pm$ 0.09 & 19.74 $\pm$ 0.16 & 19.32 $\pm$ 0.09 &            \nodata      &            \nodata      &            \nodata  \\
  $J0921+0007$ & 20.90 $\pm$ 0.26 &            \nodata      &            \nodata      & 21.05 $\pm$ 0.15 &          \nodata     & 20.38 $\pm$ 0.13  \\
  $J1406-0116$  &            \nodata      &            \nodata      &            \nodata      & 22.06 $\pm$ 0.24 &          \nodata     &          \nodata  \\
  $J0858+0000$ &            \nodata      &            \nodata      &            \nodata      & 21.19 $\pm$ 0.20 & 21.15 $\pm$ 0.20 & 21.18 $\pm$ 0.24  \\
\hline\multicolumn{7}{c}{Cool Dwarfs}\\\hline
  $J0837-0000$  &            \nodata      &            \nodata      &             \nodata     & 17.87 $\pm$ 0.01 & 17.63 $\pm$ 0.01 & 17.72 $\pm$ 0.01  \\
  $J0906-0206$  &            \nodata      &            \nodata      & 20.35 $\pm$ 0.22 &            \nodata      &            \nodata     &            \nodata  \\
  $J0912-0121$  &            \nodata      &            \nodata      &            \nodata      &          \nodata     &          \nodata    & 21.55 $\pm$ 0.31  \\
  $J1415-0113$  & 19.87 $\pm$ 0.16 & 19.67 $\pm$ 0.13 & 20.11 $\pm$ 0.21 & 20.07 $\pm$ 0.08 &            \nodata      & 19.71 $\pm$ 0.08  \\
  $J1434-0204$  & 20.17 $\pm$ 0.19 &            \nodata      &            \nodata      & 20.44 $\pm$ 0.06 & 20.18 $\pm$ 0.10 & 20.19 $\pm$ 0.08  \\
  $J1435+0040$ & 19.75 $\pm$ 0.11 & 19.84 $\pm$ 0.13 & 19.86 $\pm$ 0.15 & 19.85 $\pm$ 0.03 & 19.68 $\pm$ 0.06 & 19.87 $\pm$ 0.06  \\
\enddata
\end{deluxetable*}

\clearpage

\startlongtable
\begin{deluxetable*}{ccccccc}
\tablecaption{Spectroscopic Properties \label{tab:spectroscopy}}
\tablehead{
\colhead{Name} & \colhead{Redshift} & 
\colhead{$M_{1450}$} & \colhead{Line} & 
\colhead{EW$_{\rm rest}$} & \colhead{FWHM} & 
\colhead{log $L$}\\
\colhead{} & \colhead{} & 
\colhead{(mag)} & \colhead{} & 
\colhead{(\AA)} & \colhead{(km s$^{-1}$)} & 
\colhead{($L$ in erg s$^{-1}$)}
} 
\startdata
\multicolumn{7}{c}{Quasars}\\\hline
  $J2210+0304$ &   6.9 &   $-24.44 \pm 0.06$              & \nodata &               \nodata &               \nodata &               \nodata \\ 
  $J0213-0626$ &  6.72 &   $-25.24 \pm 0.02$           & Ly$\alpha$ &     16 $\pm$ 1 &     1200 $\pm$ 100 &    44.18 $\pm$ 0.02 \\  
  $J0923+0402$ &   6.6 &   $-26.18 \pm 0.14$              & \nodata &               \nodata &               \nodata &               \nodata \\  
  $J0921+0007$ &  6.56 &   $-24.79 \pm 0.10$           & Ly$\alpha$ &     170 $\pm$ 20 &     1400 $\pm$ 100 &    45.04 $\pm$ 0.01 \\  
  $J1545+4232$ &  6.50 &   $-24.15 \pm 0.21$           & Ly$\alpha$ &     140 $\pm$ 30 &     960 $\pm$ 660 &    44.68 $\pm$ 0.03 \\  
  $J1004+0239$ &  6.41 &   $-24.52 \pm 0.03$           & Ly$\alpha$ &     37 $\pm$ 2 &     2100 $\pm$ 300 &    44.27 $\pm$ 0.01 \\   
  $J0211-0203$ &  6.37 &   $-23.36 \pm 0.06$           & Ly$\alpha$ &     26 $\pm$ 3 &     1200 $\pm$ 900 &    43.64 $\pm$ 0.04 \\  
  $J2304+0045$ &  6.36 &   $-24.28 \pm 0.03$           & Ly$\alpha$ &     15 $\pm$ 1 &     710 $\pm$ 40 &    43.79 $\pm$ 0.03 \\  
  $J2255+0251$ &  6.34 &   $-23.87 \pm 0.04$           & Ly$\alpha$ &     19 $\pm$ 2 &     1600 $\pm$ 300 &    43.72 $\pm$ 0.04 \\  
  $J1406-0116$ &  6.33 &   $-24.96 \pm 0.06$              & \nodata &               \nodata &               \nodata &               \nodata \\  
  $J1146-0005$ &  6.30 &   $-21.46 \pm 0.63$           & Ly$\alpha$ &     260 $\pm$ 50 &     330 $\pm$ 100 &    44.18 $\pm$ 0.02 \\  
  $J1146+0124$ &  6.27 &   $-23.71 \pm 0.07$           & Ly$\alpha$ &     57 $\pm$ 6 &     9700 $\pm$ 3700 &    44.12 $\pm$ 0.03 \\  
  $J0918+0139$ &  6.19 &   $-23.71 \pm 0.04$           & Ly$\alpha$ &     23 $\pm$ 2 &     6200 $\pm$ 1700 &    43.74 $\pm$ 0.04 \\  
  $J0844-0132$ &  6.18 &   $-23.97 \pm 0.11$           & Ly$\alpha$ &     57 $\pm$ 6 &     1600 $\pm$ 300 &    44.25 $\pm$ 0.02 \\  
  $J1146-0154$ &  6.16 &   $-23.43 \pm 0.07$           & Ly$\alpha$ &     7.3 $\pm$ 4.0 &     1600 $\pm$ 400 &    43.13 $\pm$ 0.23 \\  
  $J0834+0211$ &  6.15 &   $-24.05 \pm 0.09$           & Ly$\alpha$ &     16 $\pm$ 4 &     4900 $\pm$ 800 &    43.72 $\pm$ 0.10 \\  
  $J0909+0440$ &  6.15 &   $-24.88 \pm 0.02$              & \nodata &               \nodata &               \nodata &               \nodata \\  
  $J2252+0225$ &  6.12 &   $-22.74 \pm 0.06$           & Ly$\alpha$ &     47 $\pm$ 5 &     2200 $\pm$ 600 &    43.65 $\pm$ 0.04 \\  
  $J1406-0144$ &  6.10 &   $-23.37 \pm 0.16$           & Ly$\alpha$ &     68 $\pm$ 8 &     1600 $\pm$ 500 &    44.00 $\pm$ 0.03 \\  
  $J1416+0147$ &  6.07 &   $-23.27 \pm 0.10$           & Ly$\alpha$ &     68 $\pm$ 7 &     2300 $\pm$ 200 &    44.03 $\pm$ 0.03 \\   
  $J0957+0053$ &  6.05 &   $-22.98 \pm 0.04$           & Ly$\alpha$ &     26 $\pm$ 4 &     2200 $\pm$ 1500 &    43.48 $\pm$ 0.06 \\   
  $J2223+0326$ &  6.05 &   $-25.20 \pm 0.02$           & Ly$\alpha$ &     37 $\pm$ 2 &     4700 $\pm$ 3500 &    44.53 $\pm$ 0.02 \\  
  $J1400-0125$ &  6.04 &   $-23.70 \pm 0.05$           & Ly$\alpha$ &     28 $\pm$ 4 &     12000 $\pm$ 4000 &    43.80 $\pm$ 0.05 \\  
  $J1400-0011$ &  6.04 &   $-22.95 \pm 0.11$           & Ly$\alpha$ &     56 $\pm$ 3 &     620 $\pm$ 60 &            43.84 $\pm$ 0.01 \\  
  $J1219+0050$ &  6.01 &   $-23.85 \pm 0.05$           & Ly$\alpha$ &     26 $\pm$ 4 &     7500 $\pm$ 4400 &    43.84 $\pm$ 0.06 \\  
  $J0858+0000$ &  5.99 &   $-25.28 \pm 0.01$           & Ly$\alpha$ &     24 $\pm$ 1 &     11000 $\pm$ 1000 &    44.38 $\pm$ 0.01 \\  
  $J0220-0432$ &  5.90 &   $-22.17 \pm 0.10$           & Ly$\alpha$ &     29 $\pm$ 2 &             $<$ 230 &            43.15 $\pm$ 0.03 \\ 
  $J1422+0011$ &  5.89 &   $-22.79 \pm 0.07$           & Ly$\alpha$ &     7.2 $\pm$ 1.7 &     980 $\pm$ 360 &    42.82 $\pm$ 0.10 \\  
  $J2231-0035$ &  5.87 &   $-22.67 \pm 0.10$           & Ly$\alpha$ &     21 $\pm$ 3 &     790 $\pm$ 250 &    43.23 $\pm$ 0.05 \\ 
  $J1209-0006$ &  5.86 &   $-22.51 \pm 0.05$           & Ly$\alpha$ &     26 $\pm$ 5 &     580 $\pm$ 50 &    43.01 $\pm$ 0.04 \\  
  $J1550+4318$ &  5.84 &   $-22.86 \pm 0.04$           & Ly$\alpha$ &     4.8 $\pm$ 1.5 &     2600 $\pm$ 100 &    42.70 $\pm$ 0.13 \\  
\hline\multicolumn{7}{c}{Galaxies}\\\hline
  $J1428+0159$ &  6.02 &   $-24.30 \pm 0.07$              & \nodata &               \nodata &               \nodata &               \nodata \\   
  $J0917-0056$ &  5.97 &   $-23.60 \pm 0.07$              & \nodata &               \nodata &               \nodata &               \nodata \\  
  $J0212-0315$ &  5.95 &   $-22.85 \pm 0.06$           & Ly$\alpha$  &     5.0 $\pm$ 1.0 &     650 $\pm$ 50 &    42.69 $\pm$ 0.09 \\  
  $J0212-0532$ &  5.95 &   $-22.42 \pm 0.07$              & \nodata &               \nodata &               \nodata &               \nodata \\ 
  $J2311-0050$ &  5.92 &   $-22.72 \pm 0.09$           & Ly$\alpha$  &     10 $\pm$ 1     &     250 $\pm$ 90 &    42.89 $\pm$ 0.04 \\ 
  $J1609+5515$ &  5.87 &   $-22.41 \pm 0.04$           & Ly$\alpha$ &     9.1 $\pm$ 0.6 &     480 $\pm$ 260 &  42.81 $\pm$ 0.03 \\   
  $J1006+0300$ &  5.85 &   $-22.98 \pm 0.05$              & \nodata &               \nodata &               \nodata &               \nodata \\   
  $J0914+0442$ &  5.84 &   $-23.79 \pm 0.04$              & \nodata &               \nodata &               \nodata &               \nodata \\  
  $J0219-0132$ &   5.8 &   $-22.25 \pm 0.12$              & \nodata &               \nodata &               \nodata &               \nodata \\ 
  $J0915-0051$ &  5.65 &   $-22.60 \pm 0.03$              & \nodata &               \nodata &               \nodata &               \nodata \\  
\hline\multicolumn{7}{c}{[\ion{O}{3}] Emitters}\\\hline
  $J1000+0211$ & 0.828 &               \nodata            & H$\gamma$ &     370 $\pm$ 120 &             $<$ 230 &    41.67 $\pm$ 0.03 \\   
               &       &               \nodata                                 & H$\beta$ &     860 $\pm$ 290 &             $<$ 230 &    42.04 $\pm$ 0.03 \\   
               &       &               \nodata              & [OIII] $\lambda$4959 &     2000 $\pm$ 700 &             $<$ 230 &    42.41 $\pm$ 0.01 \\   
               &       &               \nodata              & [OIII] $\lambda$5007 &     6000 $\pm$ 2000 &           $<$ 230 &    42.88 $\pm$ 0.01 \\   
  $J0154-0116$ & 0.808 &               \nodata             & H$\gamma$ &     58 $\pm$ 14       &     410 $\pm$ 290 &    41.23 $\pm$ 0.10 \\  
               &       &               \nodata                                & H$\beta$ &     85 $\pm$ 13 &             $<$ 230           &    41.39 $\pm$ 0.05 \\  
               &       &               \nodata             & [OIII] $\lambda$4959 &     130 $\pm$ 18      &     240 $\pm$ 140 &    41.57 $\pm$ 0.04 \\  
               &       &               \nodata             & [OIII] $\lambda$5007 &     320 $\pm$ 30      &             $<$ 230 &    41.96 $\pm$ 0.02 \\  
  $J2226+0237$ & 0.805 &               \nodata           & H$\gamma$ &     120 $\pm$ 20      &     390 $\pm$ 40 &    41.50 $\pm$ 0.05 \\  
               &       &               \nodata                                & H$\beta$ &     270 $\pm$ 50      &     270 $\pm$ 50 &    41.86 $\pm$ 0.03 \\  
               &       &               \nodata             & [OIII] $\lambda$4959 &     480 $\pm$ 80      &             $<$ 230 &    42.11 $\pm$ 0.01 \\  
               &       &               \nodata             & [OIII] $\lambda$5007 &     1400 $\pm$ 200  &             $<$ 230 &    42.59 $\pm$ 0.01 \\  
  $J0845-0123$ & 0.728 &               \nodata            & H$\gamma$ &               \nodata &               \nodata &               \nodata \\  
               &       &               \nodata                                & H$\beta$ &     680 $\pm$ 70     &             $<$ 230 &    42.52 $\pm$ 0.01 \\  
               &       &               \nodata             & [OIII] $\lambda$4959 &     1600 $\pm$ 200 &             $<$ 230 &    42.89 $\pm$ 0.01 \\  
               &       &               \nodata             & [OIII] $\lambda$5007 &     4700 $\pm$ 500 &             $<$ 230 &    43.37 $\pm$ 0.01 \\  
\enddata
\tablecomments{Redshifts are recorded to two significant figures when the position of Ly$\alpha$ emission or interstellar absorption is unambiguous.}
\end{deluxetable*}

\begin{deluxetable}{cc}
\tablecaption{Spectral classes of the cool dwarfs\label{tab:bdtypes}}
\tablehead{
\colhead{Name} & \colhead{Class}
} 
\startdata
  $J0158-0033$ & M5  \\
  $J0207-0052$ & T2   \\
  $J0213-0334$ & T0   \\
  $J0216-0207$ & L2   \\
  $J0220-0134$ & L2   \\
  $J0225-0351$ & M7   \\
  $J0837-0000$ & L9   \\
  $J0856+0248$ & M9  \\
  $J0900+0424$ & L8  \\
  $J0902-0030$ & M7   \\
  $J0906-0206$ & L9   \\
  $J0906+0431$ & M9  \\
  $J0912-0121$ & M7   \\
  $J1359+0134$ & M7  \\
  $J1400+0106$ & T4  \\
  $J1415-0113$ & L8   \\
  $J1432+0045$ & M8  \\
  $J1434-0204$ & T1   \\
  $J1435+0040$ & T1  \\
  $J1607+5417$ & M4  \\
  $J1620+4438$ & M5  \\
  $J1629+4233$ & M5  \\
  $J2236+0006$ & T0  \\
  $J2239-0048$ & M6   \\
  $J2248+0103$ & L2  \\
  $J2253-0117$ & T7   \\
  $J2305-0051$ & T0   \\
  $J2315-0041$ & T1   \\
\enddata
\tablecomments{These classification should be regarded as only approximate; see text.}
\end{deluxetable}

\section{Summary \label{sec:summary}}

This paper is the fourth in a series presenting the results from the SHELLQs project, a search for low-luminosity quasars at $z \ga 6$ based on
the deep multi-band imaging data produced by the HSC-SSP survey.
We continue to use the quasar selection procedure we described in Paper I and II, and 
here report spectroscopy of additional objects that roughly double the number of identifications compared to previous papers.
Through the SHELLQs project, we have so far identified 137 extremely red HSC sources over about 650 deg$^2$, which include
64 high-$z$ quasars, 24 high-$z$ luminous galaxies, 6 [\ion{O}{3}] emitters at $z \sim 0.8$, and 43 Galactic cool dwarfs.
Our discovery now exceeds, in number, the final SDSS sample of 52 high-$z$ quasars \citep{jiang16}.
The new quasars span luminosities from $M_{1450} \sim -26$ to $-22$ mag, and continue to probe a few magnitude lower luminosities 
than have been probed by previous wide-field surveys.
Our companion paper will present the quasar luminosity function established over an unprecedentedly wide range of $M_{1450} \sim -28$ to $-21$ mag, using the SHELLQs and
other survey outcomes.

Our project will continue to identify high-$z$ quasars in the HSC data, as the SSP survey continues toward its goal of observing 1,400 deg$^2$
in the Wide layer, and 27 and 3.5 deg$^2$ in the Deep and UltraDeep layers, respectively.
At the same time, we are carrying out multi-wavelength follow-up observations of the discovered objects.
Black hole masses are being measured with \ion{Mg}{2} $\lambda$2800 lines obtained with near-IR spectrographs on Subaru, the Gemini telescopes, and 
the Very Large Telescope (M. Onoue et al., in prep.).
We are also using ALMA, in order to probe the stellar and gaseous properties of the host galaxies \citep[partly published in][]{izumi18}.
The results of these observations will be presented in forthcoming papers.

\acknowledgments

This work is based on data collected at the Subaru Telescope, which is operated by the National Astronomical Observatory of Japan (NAOJ).
We appreciate the staff members of the telescope for their support during our FOCAS observations.
The data analysis was in part carried out on the open use data analysis computer system at the Astronomy Data Center of NAOJ.

This work is also based on observations made with the Gran Telescopio Canarias (GTC), installed at the Spanish Observatorio del Roque de los Muchachos 
of the Instituto de Astrof\'{i}sica de Canarias, on the island of La Palma.
We thank Stefan Geier and other support astronomers for their help during preparation and execution of our observing program.

YM was supported by the Japan Society for the Promotion of Science (JSPS) KAKENHI Grant No. JP17H04830.

The Hyper Suprime-Cam (HSC) collaboration includes the astronomical
communities of Japan and Taiwan, and Princeton University.  The HSC
instrumentation and software were developed by NAOJ, the Kavli Institute for the
Physics and Mathematics of the Universe (Kavli IPMU), the University
of Tokyo, the High Energy Accelerator Research Organization (KEK), the
Academia Sinica Institute for Astronomy and Astrophysics in Taiwan
(ASIAA), and Princeton University.  Funding was contributed by the FIRST 
program from Japanese Cabinet Office, the Ministry of Education, Culture, 
Sports, Science and Technology (MEXT), the Japan Society for the 
Promotion of Science (JSPS),  Japan Science and Technology Agency 
(JST),  the Toray Science  Foundation, NAOJ, Kavli IPMU, KEK, ASIAA,  
and Princeton University.

This paper makes use of software developed for the Large Synoptic Survey Telescope (LSST). We thank the LSST Project for 
making their code available as free software at http://dm.lsst.org.

The Pan-STARRS1 Surveys (PS1) have been made possible through contributions of the Institute for Astronomy, the University of Hawaii, the Pan-STARRS Project Office, the Max-Planck Society and its participating institutes, the Max Planck Institute for Astronomy, Heidelberg and the Max Planck Institute for Extraterrestrial Physics, Garching, The Johns Hopkins University, Durham University, the University of Edinburgh, Queen's University Belfast, the Harvard-Smithsonian Center for Astrophysics, the Las Cumbres Observatory Global Telescope Network Incorporated, the National Central University of Taiwan, the Space Telescope Science Institute, the National Aeronautics and Space Administration under Grant No. NNX08AR22G issued through the Planetary Science Division of the NASA Science Mission Directorate, the National Science Foundation under Grant No. AST-1238877, the University of Maryland, E\"{o}tv\"{o}s Lorand University (ELTE) and the Los Alamos National Laboratory.

IRAF is distributed by the National 
Optical Astronomy Observatory, which is operated by the Association of Universities for Research in Astronomy (AURA) under a cooperative agreement 
with the National Science Foundation.

\end{document}